\documentclass{pasj00}
\draft

\begin{document}
\SetRunningHead{Imanishi et al.}{HCN(4--3)/HCO$^{+}$(4--3) in NGC 4418
and Arp 220} 
%\Received{}%{yyyy/mm/dd}
%\Accepted{}%{yyyy/mm/dd}

\title{ASTE Simultaneous HCN(4--3) and HCO$^{+}$(4--3) Observations of
the Two Luminous Infrared Galaxies NGC 4418 and Arp 220}

%%% begin:list of authors
% Do NOT capitalize all letters in "textsc".
\author{Masatoshi \textsc{Imanishi}
  \thanks{Department of Astronomy, School of Science, Graduate
University for Advanced Studies (SOKENDAI), Mitaka, Tokyo 181-8588}}
\affil{National Astronomical Observatory, 2-21-1, Osawa, Mitaka, Tokyo
181-8588, Japan}
\email{masa.imanishi@nao.ac.jp}

\author{Kouichiro \textsc{Nakanishi} %
  \thanks{Department of Astronomy, School of Science, Graduate
University for Advanced Studies (SOKENDAI), Mitaka, Tokyo 181-8588}}
\affil{National Astronomical Observatory, 2-21-1, Osawa, Mitaka, Tokyo
181-8588, Japan}

\author{Masako \textsc{Yamada}}
\affil{Institute of Astronomy and Astrophysics
Academia Sinica, P.O. Box 23-141, Taipei 10617, Taiwan, R.O.C.}

\author{Yoichi \textsc{Tamura}}
\affil{Nobeyama Radio Observatory, Minamimaki, Minamisaku, Nagano,
384-1305, Japan}

\and
\author{Kotaro {\sc Kohno}}
\affil{Institute of Astronomy, University of Tokyo, 2-21-1, Osawa, Mitaka, 
Tokyo, 181-0015, Japan}
%%% end:list of authors

%%% Please use the following style in case that sorting by 
%%% affilation is impossible. 
%
% \author{%
%   D-Firstname \textsc{D-Familyname}\altaffilmark{1}
%   E-Firstname \textsc{E-Familyname}\altaffilmark{1,2}
%   and
%   F-Firstname \textsc{F-Familyname}\altaffilmark{2}}
% \altaffiltext{1}{Address of Institute}
% \email{ddddd@xxx.xxx.xx.xx}
% \email{eeeee@xxx.xxx.xx.xx}
% \altaffiltext{2}{Address of Institute}

%% `\KeyWords{}' always has to be placed before `\maketitle'.
 %Do NOT move this preamble from here!

\maketitle

\begin{abstract}
We report the results of HCN($J$=4--3) and HCO$^{+}$($J$=4--3)
observations of two luminous infrared galaxies 
(LIRGs), NGC 4418 and Arp 220, made using the Atacama
Submillimeter Telescope Experiment (ASTE).
The ASTE wide-band correlator provided simultaneous
observations of HCN(4--3) and HCO$^{+}$(4--3) lines, and a precise
determination of their flux ratios. 
Both galaxies showed high HCN(4--3) to HCO$^{+}$(4--3)
flux ratios of $>$2, possibly due to AGN-related phenomena.  
The $J$ = 4--3 to $J$ = 1--0 transition flux ratios for HCN (HCO$^{+}$) 
are similar to those expected for fully thermalized (sub-thermally
excited) gas in both sources, in spite of HCN's higher critical density. 
If we assume collisional excitation and neglect an infrared radiative
pumping process, our non-LTE analysis suggests that HCN traces gas with
significantly higher density than HCO$^{+}$. 
In Arp 220, we separated the double-peaked HCN(4--3) emission into the
eastern and western nuclei, based on velocity information.  
We confirmed that the eastern nucleus showed a higher 
HCN(4--3) to HCN(1--0) flux ratio, and thus contained a larger amount 
of highly excited molecular gas than the western nucleus.  
\end{abstract}

\section{Introduction}

Luminous infrared galaxies (LIRGs) radiate the
bulk of their large luminosities ($L\gtrsim 10^{11}L_{\odot}$) as infrared
dust emission \citep{sam96}. 
Their large infrared luminosities indicate that (1) powerful energy
sources are present, hidden behind dust; (2) energetic radiation
from the hidden energy sources is absorbed by the surrounding dust; and 
(3) the heated dust grains re-emit this energy as infrared thermal
radiation. 
To understand this LIRG population, it is essential to 
clarify the hidden energy sources, whether starbursts (i.e., nuclear fusion
inside stars) are dominant, or active galactic nuclei (AGNs; active mass
accretion onto a central compact supermassive black hole [SMBH] with
$>$10$^{6}$M$_{\odot}$) are also energetically important.   

Unlike AGNs surrounded by a torus-shaped (toroidally-shaped) dusty medium,
which are classified optically as Seyferts \citep{vei87}, 
LIRGs have a large amount of concentrated molecular gas and dust in
their nuclei \citep{sam96}.
For this reason, the putative compact AGNs can easily be {\it buried} (i.e.,
obscured in virtually all directions), making the optical detection of
AGN signatures very difficult.   
However, investigating the energetic importance of such optically
elusive {\it buried} AGNs is crucial to understand the
true nature of the LIRG population.

A starburst (nuclear fusion) and a buried AGN (mass accretion
onto a SMBH) have very different energy-generation mechanisms.
Specifically, while UV is the predominant energetic radiation in
a starburst, an AGN emits strong X-ray emission in addition to UV.
The radiative energy generation efficiency of a nuclear fusion
reaction in a normal starburst is only $\sim$0.5\% of Mc$^{2}$ (M is the
mass of material used in the nuclear fusion reaction).  
Thus, the
emission surface brightness of a starburst region is modest and has both
observational \citep{wer76,soi00} and theoretical \citep{tho05} upper limits
($\sim$10$^{13}$L$_{\odot}$  kpc$^{-2}$). 
An AGN, however, achieves high radiative energy generation
efficiency  (6--42\% of Mc$^{2}$, where M is the mass of accreting
material; Bardeen 1970; Thorne 1974), producing a very high emission
surface brightness with $>$10$^{13}$L$_{\odot}$ kpc$^{-2}$ \citep{soi00}. 
These differences between a starburst and an AGN could create
differences in the properties of the molecular gas and dust around the
energy sources, allowing us to distinguish the hidden energy sources if,
for example, different emission-line flux ratios are observed from the
surrounding material.   

Molecular gas emission lines in the sub-millimeter and millimeter
wavelength ranges are a potentially effective tool for probing this
distinction, because the effects of dust extinction are so small in
these wavelength ranges that signatures of AGNs deeply buried in gas and
dust are detectable. 
It was found observationally that HCN($J$=1--0) emission 
(rest frequency $\nu_{\rm rest}$ = 88.632 GHz or rest wavelength 
$\lambda_{\rm rest}$ = 3.385 mm) is systematically
stronger, relative to HCO$^{+}$($J$=1--0) ($\nu_{\rm rest}$ = 89.189 GHz 
or $\lambda_{\rm rest}$ = 3.364 mm),
in AGN-dominated galaxy nuclei than in starburst galaxies
\citep{koh05,kri08}.   
The trend of a strong HCN(1--0) to HCO$^{+}$(1--0) emission ratio was 
further confirmed in luminous buried AGN candidates
\citep{ima04,ink06,in06,ima07b,ima09}.
These results demonstrated that this HCN(1--0)/HCO$^{+}$(1--0) method
functions as a diagnostic tool to distinguish the hidden energy sources
of dusty galaxies. 
Extension to more distant, faint LIRGs is expected to play
an important role for the comprehensive understanding of the LIRG
population in the forthcoming Atacama Large
Millimeter/submillimeter Array (ALMA) era.
However, during the first operation phase of ALMA, band 3 
(84--116 GHz or 2.59--3.57 mm) is the longest available wavelength coverage, 
so that this HCN(1--0)/HCO$^{+}$(1--0) method can be applied only to
LIRGs at $z \leq$ 0.05, hampering further systematic investigation of LIRGs. 
Energy diagnostic methods using other transition lines and/or other
molecular species must be exploited.  

HCN($J$=4--3) ($\nu_{\rm rest}$ = 354.505 GHz) and HCO$^{+}$(4--3)
($\nu_{\rm rest}$ = 356.734 GHz) lines exist in Earth's 350 GHz
atmospheric window, and can be one of the main frequency ranges
exploited with ALMA. 
Because the frequency coverage of ALMA band 7 is 275--373 GHz, HCN(4--3) 
and HCO$^{+}$(4--3) lines can be studied in LIRGs at $z >$ 0.05.
Thus it is important to investigate whether AGNs show high
HCN(4--3) to HCO$^{+}$(4--3) flux ratios, similar to the high
HCN(1--0) to HCO$^{+}$(1--0) flux ratios.

In this paper, we present the results of simultaneous HCN(4--3) and
HCO$^{+}$(4--3) observations of two nearby, well-studied LIRGs, NGC
4418 and Arp 220, both of which show high HCN(1--0) to HCO$^{+}$(1--0)
flux ratios \citep{ima04,ima07b}. 
Throughout this paper, we adopt H$_{0}$ $=$ 75 km s$^{-1}$ Mpc$^{-1}$, 
$\Omega_{\rm M}$ = 0.3, and $\Omega_{\rm \Lambda}$ = 0.7, to be
consistent with our previously published papers.  

\section{Targets}

NGC 4418, at $z$ = 0.007, is a nearby LIRG with an infrared luminosity of
L$_{\rm IR}$ $\sim$ 9 $\times$ 10$^{10}$L$_{\odot}$ (Table 1).  
Although it shows no clear Seyfert signature in the optical spectrum
\citep{arm89,leh95}, it is considered one of the closest and strongest
buried AGN candidates, based on the AGN-like infrared spectral shape
\citep{roc91,dud97,spo01} and the high HCN(1--0) to HCO$^{+}$(1--0) flux
ratio in the millimeter wavelength range \citep{ima04}.

Arp 220 (z = 0.018) is one of the best-studied nearby ultraluminous 
infrared galaxies (L$_{\rm IR}$ $>$ 10$^{12}$L$_{\odot}$; Sanders
et al. 1988). 
It has two nuclei, the eastern (Arp 220 E) and  western (Arp 220 W)
nuclei, separated by $\sim$1$''$ \citep{sco00,soi00}.
The combined optical spectrum of both nuclei is classified as a LINER
(i.e., non-Seyfert) \citep{vei99}.
Starburst activity has been detected, but was
energetically insufficient to account quantitatively for 
the observed infrared luminosity \citep{idm06,arm07,ima07a},
indicating an energy source deeply buried in Arp 220's nuclear core
\citep{spo04,gon04}.   
The energy source in Arp 220 W has a very high emission
surface brightness, for which a buried AGN is a plausible (if not 
definite) explanation \citep{dow07,sak08}. 

\section{Observations and Data Reduction}

The HCN(4--3) and HCO$^{+}$(4--3) observations of NGC 4418 and Arp 220
were made using the Wideband and High dispersion Spectrometer system
with FFX correlator (WHSF) \citep{oku08,igu08} on the Atacama
Submillimeter Telescope Experiment (ASTE), a 10-m antenna located at
Pampa La Bola in the 
Atacama Desert of Chile, at an altitude of 4800 m \citep{eza04,eza08}. 
Because the WHSF has 4-GHz frequency coverage in each side band, both the
HCN(4--3) ($\nu_{\rm rest}$ = 354.505 GHz) and HCO$^{+}$(4--3)
($\nu_{\rm rest}$ = 356.734 GHz) lines were simultaneously observable.  
Figure 1 shows the raw ASTE WHSF spectra of NGC 4418 and Arp 220.

The observations were made in 2008 May, remotely from an ASTE operation
room in the Mitaka campus of the National Astronomical Observatory of
Japan, using the network observation system N-COSMOS3 developed by
\citet{kam05}.   
Table 2 summarizes the detailed observation log.
The weather conditions were excellent. 
Typical system temperatures were 180--300 K in the single side band.  
We employed the position-switching mode, so that the target position and
off-sky position, 4 arcmin away from the target, were switched after
each 10 s exposure. 
Telescope pointing was checked every 2 h. 
The total net on-source integration times were 200 min and 155 min for
NGC 4418 and Arp 220, respectively. 
Flux calibration was made every observation day. 
By changing the frequency setting of the WHSF, we observed a 
strong emission line in W28 (IRC10216) at the frequency of the red-shifted 
HCN(4--3) and HCO$^{+}$(4--3) lines of NGC 4418 (Arp 220), to
estimate the possible sensitivity variation along the WHSF bandpass.
This variation was found to be $<$20\%. 
The half-power beam width of the ASTE 10-m dish was 22$''$
at $\nu$ $\sim$ 345 GHz. 
At this frequency, the main beam efficiency is 0.6, and 1 K in antenna 
temperature corresponds to 78.5 Jy. 

Data reduction was carried out with NEWSTAR, a package developed at
Nobeyama Radio Observatory.  
A fraction of poor-quality data was flagged. 
To estimate the strength of HCN(4--3) and HCO$^{+}$(4--3) emission, 
we subtracted baselines from the obtained raw spectra.
We first divided the spectra into two parts at the center, the
lower-frequency HCN(4--3) part and higher-frequency HCO$^{+}$(4--3)
part. 
The baseline was estimated independently in each part, based on the 
data points free from emission lines. 

For NGC 4418, ten data sets with 20 min net on-source exposure time for
each were obtained. 
The behavior of the baseline was extremely good, and the HCN(4--3)
emission line was easily recognizable in individual data sets.
First-order linear baselines were adopted both for the HCN(4--3)
and HCO$^{+}$(4--3) emission lines, and the baseline-subtracted
individual spectra were summed to obtain final spectra.  

For Arp 220, final data consist of seven data sets with 20 min net
on-source exposure and one data set with 15 min net on-source exposure.  
For the HCN(4--3) line, first-order linear baselines were adopted 
in individual data sets, and the baseline-subtracted individual
spectra were summed.  
The HCN(4--3) emission line was detected in the final spectrum (see
Appendix for more detail).  
Although we tried several baseline fits, the HCN(4--3)
emission line was always visible, and its profile was virtually
unchanged in the final spectrum, under reasonable baseline choices.
For the HCO$^{+}$(4--3) line, when we applied a first-order
linear baseline, winding patterns remained in the final
spectrum, and HCO$^{+}$(4--3) line was not clearly detected. 
We thus attempted a second-order baseline, but the
HCO$^{+}$(4--3) emission line remained undetected in the final spectrum. 

\section{Results}

Figure 2 presents the final spectra around the HCN(4--3) and
HCO$^{+}$(4--3) emission lines, after baseline subtraction. 
In both NGC 4418 and Arp 220, HCN(4--3) emission was clearly detected, 
with peak antenna temperatures of 7--15 mK, but HCO$^{+}$(4--3)
emission was not.    
For NGC 4418, HCN(4--3) and HCO$^{+}$(4--3) emission line spectra are 
published here for the first time.  

For NGC 4418, we fit the detected HCN(4--3) emission with a single
Gaussian profile.
For Arp 220 HCN(4--3) emission, a double-peaked profile was clearly
seen. Thus we fit it with a two-component Gaussian profile.
Table 3 summarizes the fitting results.
The estimated HCN(4--3) fluxes were 220 Jy km s$^{-1}$ and
410 Jy km s$^{-1}$ for NGC 4418 and Arp 220, respectively, with the
uncertainties of $\sim$20\%.
For the undetected HCO$^{+}$(4--3) emission, we adopted the lowest
plausible continuum level in the spectra, and estimated conservative
upper limits, by counting the possible flux excess above the adopted
continuum levels. 
Table 3 summarizes these upper limits for the HCO$^{+}$(4--3) emission.

For Arp 220, HCN(4--3) measurements have been made previously by two
groups \citep{wie02,gre09}.   
Our ASTE measurement of the HCN(4--3) flux (410$\pm$82 Jy km s$^{-1}$)
fell between the previously reported values of 260 Jy km s$^{-1}$ 
\citep{wie02} and 587$\pm$118 Jy km s$^{-1}$ \citep{gre09}.
The velocity difference in the two components in the double-peaked HCN(4--3)
profile was $\sim$300 km s$^{-1}$ (Table 3), similar to that of
HCN(1--0) emission \citep{ima07b}, further strengthening the detection
of the HCN(4--3) emission line in Arp 220.  
Based on the argument of \citet{ima07b}, we ascribe the lower-velocity 
(blue-shifted) and higher-velocity (red-shifted) components to the Arp 220
W and E nuclei, respectively.

The HCN(4--3) to HCO$^{+}$(4--3) flux ratios
%------------
\footnote{By definition, flux is proportional to brightness-temperature
$\times$ $\nu^{2}$, where $\nu$ is frequency.  
Because HCN(4--3) ($\nu_{\rm rest}$ = 354.505 GHz) and HCO$^{+}$(4--3)
($\nu_{\rm rest}$ = 356.734 GHz) have very similar frequencies, their 
flux ratios are virtually identical to their brightness-temperature
ratios. 
In this paper, we use a flux ratio.
}
%------------
were $>$2.7 and $>$3.0 for NGC 4418 and Arp 220, respectively, where the
ratio for Arp 220 used the combined emission from both nuclei.  
The non-detection of HCO$^{+}$(4--3) emission lines hampered the 
spectral separation into Arp 220 E and W and rendered independent
estimates of the line flux ratio for these nuclear components difficult.
HCN(4--3)/HCO$^{+}$(4--3) flux ratios significantly higher than unity
were confirmed in both galaxies.  

We emphasize that our simultaneous HCN(4--3) and HCO$^{+}$(4--3)
observations using ASTE WHSF are crucial for obtaining reliable
HCN(4--3) to HCO$^{+}$(4--3) flux ratios.  
Submillimeter and millimeter observations still present
non-negligible uncertainties in absolute flux calibrations.  
In fact, even for measurements obtained in the last several years, the
HCN(4--3) and HCN(1--0) fluxes of Arp 220 differed by a factor of 2 at
maximum as observed by different groups with different telescopes under
different weather conditions and using different calibration methods 
\citep{wie02,ima07b,gra08,gre09}. 
This level of absolute flux calibration uncertainty can be fatal to an 
investigation of the HCN(4--3) to HCO$^{+}$(4--3) flux ratios because 
we are discussing the difference of the ratios by about a factor of 2.
By simultaneously observing HCN(4--3) and HCO$^{+}$(4--3) lines, the
possible absolute flux calibration uncertainty does not propagate to the 
HCN(4--3)/HCO$^{+}$(4--3) flux ratios, which are then dominated only by
statistical noises and fitting errors.  

\section{Discussion}

\subsection{HCN(4--3) to HCO$^{+}$(4--3) flux ratios}

In this subsection, we explore HCN(4--3) and HCO$^{+}$(4--3) emission
from NGC 4418 and Arp 220, in relation to HCN(1--0) and HCO$^{+}$(1--0).  
For this purpose, we used our ASTE WHSF and Nobeyama Millimeter
Array (NMA) data \citep{ima04,ima07b}, because (1) simultaneous 
HCN and HCO$^{+}$ observations by both telescopes make the comparison of HCN
to HCO$^{+}$ flux ratios reliable, and (2) HCN(1--0) and
HCO$^{+}$(1--0) absolute fluxes in the NMA data have been confirmed to 
agree with measurements
from other single-dish telescopes \citep{ima04,ima07b}.  
The NMA measurements of HCN(1--0) and HCO$^{+}$(1--0) fluxes are summarized
in Table 4.

For NGC 4418, we obtained an HCN(4--3) to HCO$^{+}$(4--3) flux ratio 
$>$ 2.7, confirming the high HCN to HCO$^{+}$ flux ratio previously 
found for the $J$ = 1--0 transition line ($\sim$1.8; Imanishi et al. 2004).
The HCN(4--3) to HCN(1--0) flux ratio was estimated to be 20.  
This is comparable to the $\nu^{2}$ scaling ($\sim$16) expected for  
fully thermalized optically thick gas \citep{rie06}.
The HCO$^{+}$(4--3) to HCO$^{+}$(1--0) flux ratio was $<$15.

For Arp 220, the HCN(4--3) and HCO$^{+}$(4--3) flux ratio was estimated
to be $>$3.0.  
The $J$ = 4--3 to $J$ = 1--0 flux ratio for Arp 220 total emission was 12
and $<$8 for HCN and HCO$^{+}$, respectively.
When we separated the emission from individual nuclei, the HCN(4--3) to
HCN(1--0) flux ratios were 21 and 9 for Arp 220 E and W nuclei,
respectively. 
The ratio is higher in Arp 220 E than Arp 220 W, further supporting the 
previous argument that a larger fraction of highly excited molecular gas
is present in Arp 220 E \citep{gre09}. 

The critical density of HCN is 2.3 $\times$ 10$^{5}$ cm$^{-3}$ 
for $J$ = 1--0 and 8.5 $\times$ 10$^{6}$ cm$^{-3}$ for $J$ = 4--3.
That of HCO$^{+}$ is 3.4 $\times$ 10$^{4}$ cm$^{-3}$ for $J$ = 1--0, and 
1.8 $\times$ 10$^{6}$ cm$^{-3}$ for $J$ = 4--3 (Table 3 of Greve et
al. 2009). 
Because the critical density of HCO$^{+}$ is a factor of $\sim$5--7 
smaller than HCN for both the $J$ = 1--0 and $J$ = 4--3 transitions,
HCO$^{+}$ can be collisionally excited more easily than HCN in
single-phase molecular gas. 
Thus, the stronger HCN emission in NGC 4418 and Arp 220 requires 
(1) an HCN abundance that is significantly higher than HCO$^{+}$, and/or 
(2) the presence of some physical mechanism that enhances HCN emission.

Regarding the first HCN abundance enhancement scenario,
\citet{yam07} concluded, based on three-dimensional line transfer
simulations, that HCN abundance must be an order 
of magnitude higher than HCO$^{+}$ to account for the observed high
HCN to HCO$^{+}$ flux ratio ($\sim$2).
While the typical HCN to HCO$^{+}$ abundance ratio in our Galactic
molecular cloud is $\sim$1 \citep{bla87,pra97,dic00}, 
HCN overabundance, relative to HCO$^{+}$, by an order of magnitude, is
predicted in some parameter range for dense molecular gas illuminated
by an X-ray emitting energy source \citep{mei05}.
Since an AGN is a much stronger X-ray emitter than starburt activity, 
this HCN overabundance scenario is a possibility for the observed
high HCN(4--3)/HCO$^{+}$(4--3) flux ratios in the buried AGN candidates,
NGC 4418 and Arp 220. 

Strong HCN emission in luminous buried AGN candidates may be explained 
by the infrared radiative pumping scenario
\citep{aal95,gar06,gue07,wei07,aal07a}.  
In an AGN, the emission surface brightness is much higher than
in a starburst, so that the surrounding dust is heated to a high temperature
(several hundred K) and produces strong mid-infrared 10--20 $\mu$m continuum
emission. 
The HCN molecule has a line at infrared 14.0 $\mu$m.
Molecular gas around an AGN can be vibrationally excited by absorbing 
these infrared 14.0 $\mu$m photons, and through the subsequent cascade
process, can enhance HCN rotational lines in the sub-millimeter and
millimeter range.   

\citet{gue07} suggested that this infrared pumping scenario may work also 
for HCO$^{+}$ and HNC, because they have lines at 12.1 $\mu$m
(HCO$^{+}$) and 21.7 $\mu$m (HNC).
However, for this infrared pumping scenario to affect
particular molecules, infrared photons at the wavelengths of these
molecular lines must be absorbed. 
Namely, absorption features must be detected in infrared
spectra at 14.0 $\mu$m (HCN), 12.1 $\mu$m (HCO$^{+}$), and 21.7 $\mu$m (HNC).
While the infrared HCN 14.0 $\mu$m absorption feature is clearly 
detected in highly obscured ultraluminous infrared galaxies, including 
NGC 4418 and Arp 220 \citep{lah07,vei09}, absorption features at
HCO$^{+}$ 12.1 $\mu$m and HNC 21.7 $\mu$m are not \citep{vei09}.
Therefore, we see observational signatures that this infrared radiative
pumping scenario could be at work for HCN, but not for HCO$^{+}$ or 
HNC. This may explain the strong HCN emission found in luminous AGN
candidates.

Since both the HCN overabundance and infrared radiative pumping
scenarios could work more effectively in an AGN than starburst activity,
the enhanced HCN(4--3) emission may originate in putative buried AGNs in
NGC 4418 and Arp 220.
Three-dimensional modelings which realistically incorporate actual
physical parameters of molecular gas around an AGN are needed for 
more detailed quantitative comparison with observations. 

\subsection{Molecular gas properties}

For NGC 4418, the HCN(4--3) to HCN(1--0) flux ratio (20) was higher than the
HCO$^{+}$(4--3) to HCO$^{+}$(1--0) flux ratio ($<$15).  
For Arp 220 as well, the former ratio (12) was higher than 
the latter ratio ($<$8).
Thus, our ASTE observations suggest that $J$ = 4--3 transition lines are
excited more efficiently in HCN than HCO$^{+}$, in spite of the
higher critical density of HCN(4--3) compared to HCO$^{+}$(4--3) (a
factor of $\sim$5 [$\S$5.1]).
As mentioned in $\S$5.1, in a single-phase molecular gas, molecular
species with lower critical density are more easily
collisionally excited.  
Thus, HCO$^{+}$ should be excited more than HCN. 
Some mechanisms are needed to explain the strong HCN(4--3) emission. 
Examples include a multiple phase molecular gas \citep{gre09}, and  
the infrared radiative pumping scenario.
The second scenario will be investigated elsewhere, using a 
three-dimensional model (Yamada et al. 2009, in prep).
Here, we explore the first possibility. 
Namely, we consider only collisional excitation and neglect the infrared
radiative pumping process.  

We used RADEX \citep{van07} to investigate molecular gas properties that
could account for our ASTE and NMA results. 
The RADEX software adopts the non-LTE analysis of molecular line spectra,
solving radiative transfer in simple geometry gas.
We varied the H$_{2}$ number density and kinetic temperature from 
10$^{3}$--10$^{8}$ cm$^{-3}$ and 10--200 K, respectively.
A uniform sphere with a line width of 200 km s$^{-1}$ was assumed, 
based on the HCN(4--3) emission measurements (Table 3).
X-ray observations suggested that both NGC 4418 and Arp 220 have
highly-obscured, Compton thick (N$_{\rm H}$  $>$ 10$^{24}$ cm$^{-2}$)
AGNs \citep{mai03,iwa05}.  
Taking HCN and HCO$^{+}$ abundances relative to H$_{2}$ of
10$^{-9}$$-$10$^{-8}$ \citep{bla87,dic00}, column densities with  
10$^{16}$ cm$^{-2}$ were adopted for both HCN and HCO$^{+}$. 
The T = 2.73 K background emission was also added for the
calculation.
Figure 3 shows the HCN(4--3) to HCN(1--0) and 
HCO$^{+}$(4--3) to HCO$^{+}$(1--0) flux ratios in Jy km s$^{-1}$, 
as a function of molecular gas number density and kinetic temperature.
In the 10--200 K temperature range, the molecular gas number density was
estimated to be n$_{\rm H}$ $>$10$^{6}$ cm$^{-3}$ to fit the 
observed HCN(4--3) to HCN(1--0) flux ratios for NGC 4418 ($\sim$20) and
Arp 220 ($\sim$12).

Now, the HCO$^{+}$(4--3) to HCO$^{+}$(1--0) flux ratios were 
$<$15 and $<$8 for NGC 4418 and Arp 220, respectively.
For T $>$ 25 K ($>$10$^{1.4}$ K),
molecular gas with a number density n$_{\rm H}$ $<$10$^{6}$ cm$^{-3}$
reproduced the observed HCO$^{+}$ flux ratios.
Thus, assuming collisional excitation, we confirmed that in both NGC
4418 and Arp 220, HCN emission traces gas with a number density
significantly higher than that traced by HCO$^{+}$ 
%------------
\footnote{For NGC 4418, the small HCN(3--2) to HCO$^{+}$(3--2) flux
ratio (less than unity) reported by \citet{aal07b} is not compatible
with our model. 
The HCN(3--2) flux itself is also significantly lower than that expected
for thermalized, optically thick gas. We have no clear interpretation,
as long as only collisional excitation is considered.
},
%------------
as previously argued for Arp 220 based on the measurements of
HCN(4--3), HCO$^{+}$(4--3), and other transition lines \citep{gre09}.
Even given HCN and HCO$^{+}$ column densities of  
10$^{15}$ cm$^{-2}$ and a line width of 300 km s$^{-1}$, 
our main conclusion is unchanged.

Three-dimensional hydrodynamic simulations of galactic disks have shown
that the fraction of molecular gas with n$_{\rm H}$ $>$ 10$^{6}$
cm$^{-3}$ is miniscule \citep{wad07}. 
Yet, the strong HCN(4--3) emission line detected in NGC 4418 and Arp 220
suggests the presence of a large amount of such high-density 
(n$_{\rm H}$ $>$ 10$^{6}$ cm$^{-3}$) molecular gas, if the HCN(4--3)
line is collisionally excited. 
It is likely that the high nuclear concentration of molecular gas in LIRGs 
increases the fraction of high-density molecular gas. 

\section{Summary}

We presented the results of simultaneous HCN(4--3) and HCO$^{+}$(4--3)
observations of two luminous infrared galaxies, NGC 4418 and Arp 220,
using ASTE WHSF. Our main conclusions are as follows:

\begin{enumerate}

\item Strong HCN(4--3) emission was detected in both NGC 4418 and Arp 220,
but HCO$^{+}$(4--3) emission provided only upper limits.

\item In both sources, we found HCN(4--3) to HCO$^{+}$(4--3)
flux ratios of $>$2, even higher than the HCN(1--0)
to HCO$^{+}$(1--0) emission line flux ratios. 
The high HCN/HCO$^{+}$ flux ratios could be explained by
HCN overabundance and/or infrared radiative pumping scenarios, both of
which may work more effectively in an AGN than starburst activity. 

\item HCN(4--3) to HCN(1--0) flux ratios were higher than
HCO$^{+}$(4--3) to HCO$^{+}$(1--0) flux ratios in both NGC 4418 and Arp
220, suggesting that higher $J$-transitions were excited more
efficiently in HCN than HCO$^{+}$. 
If we neglect an infrared radiative pumping mechanism, and consider only
collisional excitation, these results are incompatible with a one-zone
molecular gas model, because the critical density of HCN is higher than
HCO$^{+}$ at these transitions. A multi-phase molecular gas in which HCN
selectively probes higher-density molecular gas than HCO$^{+}$ is required.
We applied RADEX models, and found that in both NGC 4418 and Arp 220, gas
with a number density n$_{\rm H}$ $>$ 10$^{6}$ cm$^{-3}$ and 
n$_{\rm H}$ $<$ 10$^{6}$ cm$^{-3}$ reproduced the observed HCN and
HCO$^{+}$ $J$ = 4--3 to 1--0 flux ratios, respectively, in the 25--200 K
temperature range.  

\item For Arp 220, HCN(4--3) to HCN(1--0) flux ratio is higher in Arp
220 E than in Arp 220 W, suggesting that Arp 220 E contains a larger
amount of highly excited molecular gas.    

\end{enumerate}

\vspace{1cm}

%\acknowledgments

We thank T. Okuda for the support of our ASTE WHSF observing run, and
the anonymous referee for his/her very useful comments.
The ASTE project is driven by Nobeyama Radio Observatory (NRO), a branch
of National Astronomical Observatory of Japan (NAOJ), in collaboration
with University of Chile, and Japanese institutes including University
of Tokyo, Nagoya University, Osaka Prefecture University, Ibaraki
University, and Hokkaido University.
Observations with ASTE were in part carried out remotely from Japan by
using NTT's GEMnet2 and its partner R\&E (Research and Education)
networks, which are based on AccessNova collaboration of University of
Chile, NTT Laboratories, and NAOJ.
M.I. is supported by Grants-in-Aid for Scientific Research (19740109). 
This study utilized the NASA/IPAC Extragalactic Database (NED) operated
by the Jet Propulsion Laboratory, California Institute of Technology,
under contract with the National Aeronautics and Space Administration.

\clearpage

\appendix

\section{Analysis of the Arp 220 HCN(4--3) emission line}

We explain in detail how the HCN(4--3) emission line of Arp 220 becomes
detectable in our analysis.  
We took eight independent data sets.
Single-order linear baselines were employed in individual data sets
(Figure 4) and eight baseline-subtracted spectra were summed to obtain
a final HCN(4--3) spectrum. 
The HCN(4--3) emission was clearly detected in the final spectrum of Arp
220, as seen in Figure 2.

\clearpage

%----------------- Tables -----------------%
%---- Table 1 ----%
\begin{table}[h]
%\small
%\scriptsize
\caption{Detailed information on the observed LIRGs \label{tab1}}
\begin{center}
\begin{tabular}{lcccrrc}
\hline
\hline
Object & Redshift & $f_{\rm 12}$ & $f_{\rm 25}$ & 
$f_{\rm 60}$ & $f_{\rm 100}$ & log $L_{\rm IR}$ (log $L_{\rm IR}$/$L_{\odot}$) \\ 
 & & (Jy) & (Jy) & (Jy) & (Jy) & (ergs s$^{-1}$) \\
(1) & (2) & (3) & (4) & (5) & (6) & (7)  \\
\hline
NGC 4418 & 0.007 $^{a}$ & 0.9 & 9.3 & 40.7 & 32.8 & 44.5 (10.9) \\ 
Arp 220 & 0.018 $^{b}$ & 0.48 & 7.92 & 103.33 & 112.40 & 45.7 (12.1) \\ \hline  
\end{tabular}
\end{center}
\end{table}

Notes.

Col.(1): Object name.

Col.(2): Redshift.

Cols.(3)--(6): f$_{12}$, f$_{25}$, f$_{60}$, and f$_{100}$ are
{\it IRAS FSC} fluxes at 12, 25, 60, and 100 $\mu$m, respectively.

Col.(7): Decimal logarithm of the infrared (8$-$1000 $\mu$m) luminosity
in ergs s$^{-1}$ calculated as follows:
$L_{\rm IR} = 2.1 \times 10^{39} \times$ D(Mpc)$^{2}$
$\times$ (13.48 $\times$ $f_{12}$ + 5.16 $\times$ $f_{25}$ +
$2.58 \times f_{60} + f_{100}$) ergs s$^{-1}$
\citep{sam96}.
The values in parentheses are the decimal logarithms of the infrared
luminosities in units of solar luminosities.

$^{a}$: 1 arcsec corresponds to 130 pc.

$^{b}$: 1 arcsec corresponds to 340 pc.

%---- Table 2 ----%
\begin{table}[h]
%\small
\caption{Observing log of ASTE observations \label{tab2}}  
\begin{center}
\begin{tabular}{llccc}
\hline
\hline
Object & Observing Date & Central & Pointing & Flux \\  
 & (UT) & frequency & calibrator & calibrator \\   
(1) & (2) & (3) & (4) & (5) \\ \hline 
NGC 4418 & 2008 May 22, 23, 24, 27 & 353.15 & V Hya, IRC10216 & W28 \\   
Arp 220  & 2008 May 24, 27 & 349.21 & RX Boo & IRC10216 \\  \hline
\end{tabular}
\end{center}
\end{table}

Notes. 

Col.(1): Object name.

Col.(2): Observation date in UT.

Col.(3): Central frequency of ASTE WHSF used for the observations.

Col.(4): Pointing calibrator object name.

Col.(5): Flux calibrator object name.

\clearpage

%---- Table 3 ----%
\begin{table}
\scriptsize
\caption{Properties of HCN(4--3) and HCO$^{+}$(4--3) emission lines \label{tab3}}    
\begin{center}
\begin{tabular}{llcccccc}
\hline
\hline 
Object & Line & LSR velocity & Line width & Flux & Flux &Luminosity & Ratio \\   
 &  & [km s$^{-1}$] & [km s$^{-1}$] & [K km s$^{-1}$] & [Jy km s$^{-1}$] &
10$^{7}$ [K km s$^{-1}$ pc$^{2}$] &  \\  
(1) & (2) & (3) & (4) & (5) & (6) & (7) & (8) \\ \hline 
NGC 4418 & HCN(4--3) & 2060 & 335 & 4.7 & 220 & 4.7 & $>$2.7 \\
         & HCO$^{+}$(4--3) & --- & --- & $<$1.7 & $<$80 & $<$1.7 &  \\
Arp 220  & HCN(4--3) & 5265, 5560 & 220,200 & 8.7 & 410 & 57 & $>$3.0 \\
         & & & & 3.4 (W) + 5.3 (E) & 160 (W) + 250 (E) & 22 (W) + 35 (E)& \\
         & HCO$^{+}$(4--3) & --- & --- & $<$2.9 & $<$135 & $<$19& \\ \hline 
\end{tabular}
\end{center}
\end{table}

Notes. 

Col.(1): Object name.

Col.(2): HCN(4--3) or HCO$^{+}$(4--3) line.

Col.(3): LSR velocity \{v$_{\rm opt}$ $\equiv$ 
($\frac{\nu_0}{\nu}$ $-$ 1) $\times$ c\} of the
HCN(4--3) and HCO$^{+}$(4--3) in [km s$^{-1}$].

Col.(4): Line width in FWHM of the HCN(4--3) and HCO$^{+}$(4--3)
emission lines in [km s$^{-1}$]. 

Col.(5): Flux of the HCN(4--3) and HCO$^{+}$(4--3) emission lines in
main beam temperature in [K km s$^{-1}$]. 
For ASTE, the main beam efficiency is $\eta$ $\sim$ 0.6 at $\nu$ $\sim$ 
345 GHz.
For the HCN(4--3) line of Arp 220, individual contributions from Arp 220 E
and W are also shown. 
The flux uncertainty is estimated to be $<$20\%, based on the 
combination of the empirical one of ASTE WHSF (10--15\%) and
systematic one of flux calibrators ($\sim$10\%) \citep{wan94}. 
  
Col.(6): Flux of the HCN(4--3) and HCO$^{+}$(4--3) emission lines 
in [Jy km s$^{-1}$].
For ASTE, 1 [K] in main beam temperature corresponds to 47 [Jy] at  
$\nu$ $\sim$ 345 GHz. 

Col.(7): Luminosity of the HCN(4--3) and HCO$^{+}$(4--3) emission lines  
in $\times$10$^{7} $[K km s$^{-1}$ pc$^{2}$].

Col.(8): HCN(4--3)/HCO$^{+}$(4--3) flux ratio (flux $\propto$ $\nu^{2}$
$\times$ brightness-temperature). 
Because both the HCN(4--3) and HCO$^{+}$(4--3) data are taken simultaneously, 
the ratio is not affected by possible absolute flux calibration
uncertainties in the ASTE observations (see $\S$4).

\clearpage

%---- Table 4 ----%
\begin{table}[h]
%\small
\caption{HCN and HCO$^{+}$ data at $J$ = 4--3 and $J$ = 1--0 transition
lines for NGC 4418 and Arp 220 \label{tab3}}   
\begin{center}
\begin{tabular}{llcl}
\hline
\hline
Object & Line & Flux & Reference \\   
(1) & (2) & (3) & (4) \\ \hline
NGC 4418 & HCN(4--3) & 220$\pm$44  & This work \\ 
         & HCO$^{+}$(4--3) & $<$80 & This work \\
         & HCN(1--0) & 10.3  & \citet{ima04} \\ 
%         &           & 8.5 \tablenotemark{b} & \citet{ima04} \\ 
         & HCO$^{+}$(1--0) & 5.5    & \citet{ima04} \\  \hline
%         &                 & 11     & \citet{aal07} \\ \hline
%        & HCN(3--2) & 26 \tablenotemark{a}  & \citet{aal07} \\ 
%        & HCO$^{+}$(3--2) & 45 \tablenotemark{a} & \citet{aal07} \\
Arp 220  & HCN(4--3) & 410$\pm$82  & This work \\ 
         &           & 250 (E), 160 (W) & This work \\
         &           & 260  & \citet{wie02} \\ 
         &           & 587$\pm$118  & \citet{gre09} \\ 
         & HCO$^{+}$(4--3)  & $<$135 & This work \\
         &           & 106$\pm$23  & \citet{gre09} \\ 
         &           & 80   & \citet{sak09} \\
         & HCN(1--0) & 34  & \citet{ima07b} \\ 
         &           & 12 (E), 18 (W) & \citet{ima07b} \\
         & HCO$^{+}$(1--0) & 18  & \citet{ima07b} \\ 
         &           & 5.5 (E), 6.5 (W) & \citet{ima07b} \\ \hline
\end{tabular}
\end{center}
\end{table}

Notes. 

Col.(1): Object name.

Col.(2): HCN or HCO$^{+}$ line.

Col.(3): Flux in [Jy km s$^{-1}$].
For Arp 220, individual fluxes from Arp 220 E and W are also shown.
For HCN(1--0) and HCO$^{+}$(1--0) fluxes from Arp 220 E and W, only
the spatially unresolved core emission is extracted.

Col.(4): Reference.

\clearpage

%-----------------  Figures  ---------------% 
%---- Figure 1 ----%
\begin{figure}
\FigureFile(85mm,85mm){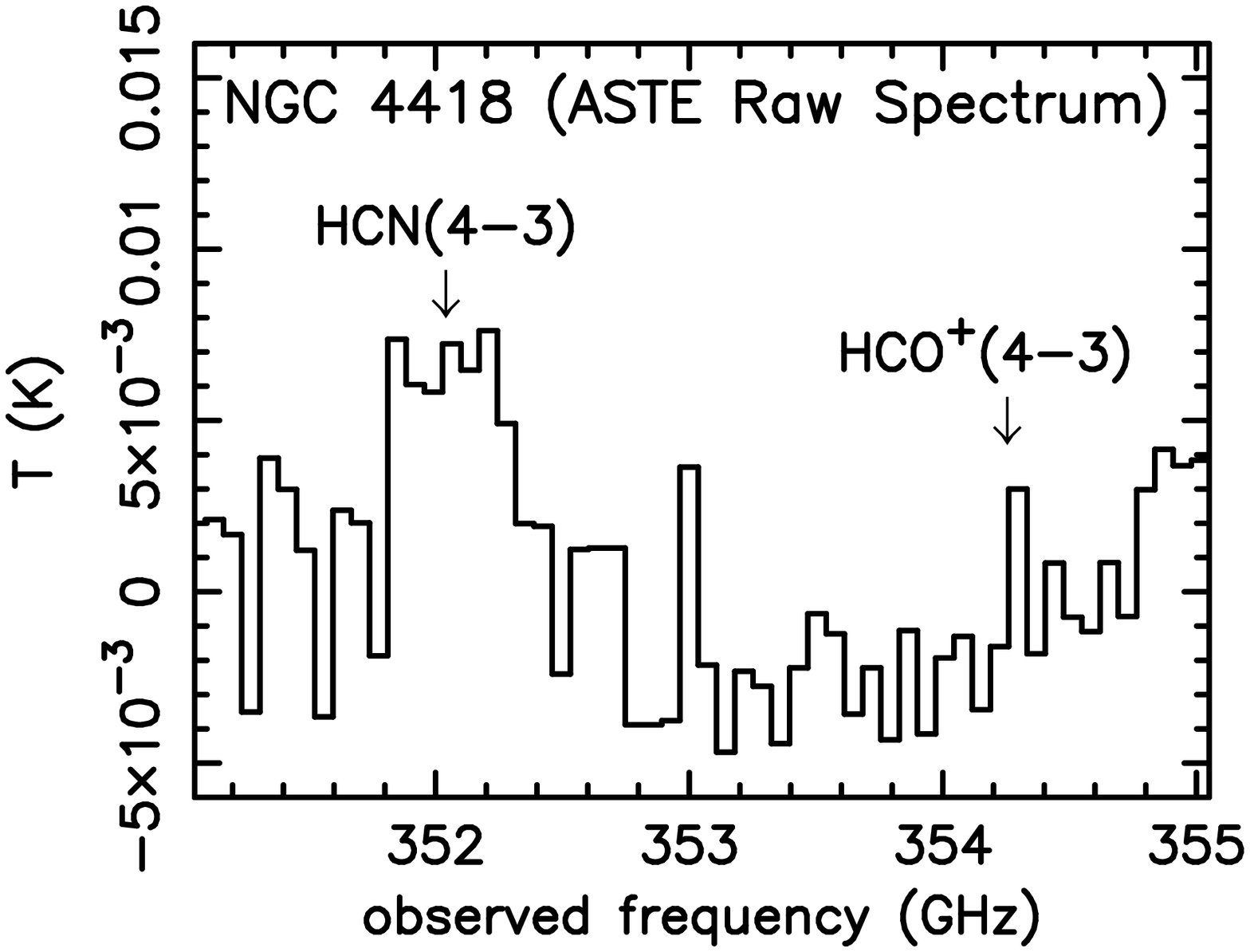} 
\FigureFile(85mm,85mm){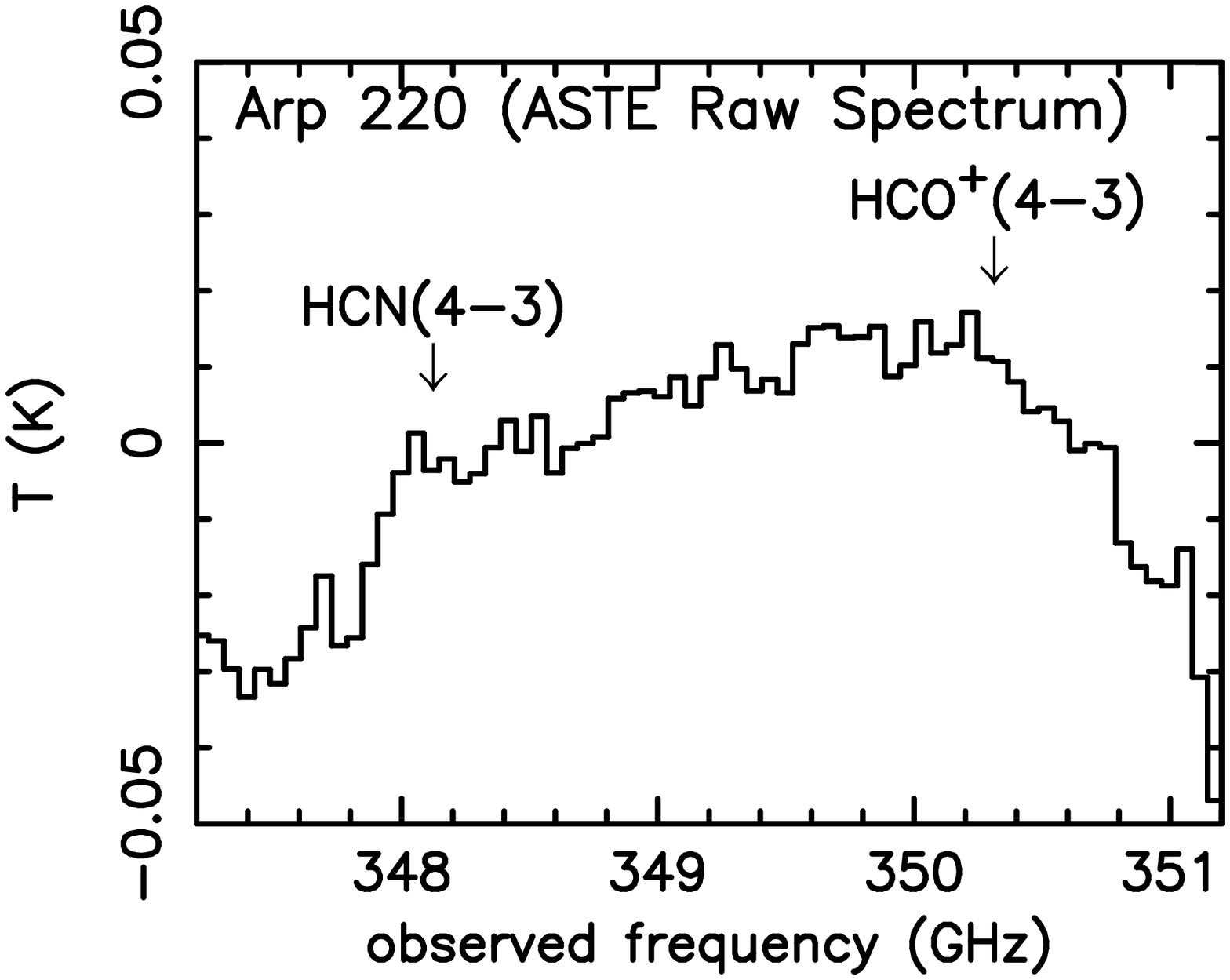} 
%\FigureFile(80mm,80mm){ps/N4418raws.ps} 
\caption{
ASTE WHSF wide-band (4 GHz) spectra of NGC 4418 and Arp 220, before
baseline subtraction. 
The abscissa is observed frequency in [GHz] and the ordinate is antenna
temperature in [K].
The spectrum is smoothed to a velocity resolution of 60 km s$^{-1}$ and 
50 km s$^{-1}$ for NGC 4418 and Arp 220, respectively. 
The expected frequency of the HCN(4--3) and HCO$^{+}$(4--3) lines are
indicated as arrows.  
HCN(4--3) and HCO$^{+}$(4--3) emission lines are
simultaneously covered. 
}
\end{figure}

%---- Figure 2 ----%
\begin{figure}
\FigureFile(85mm,85mm){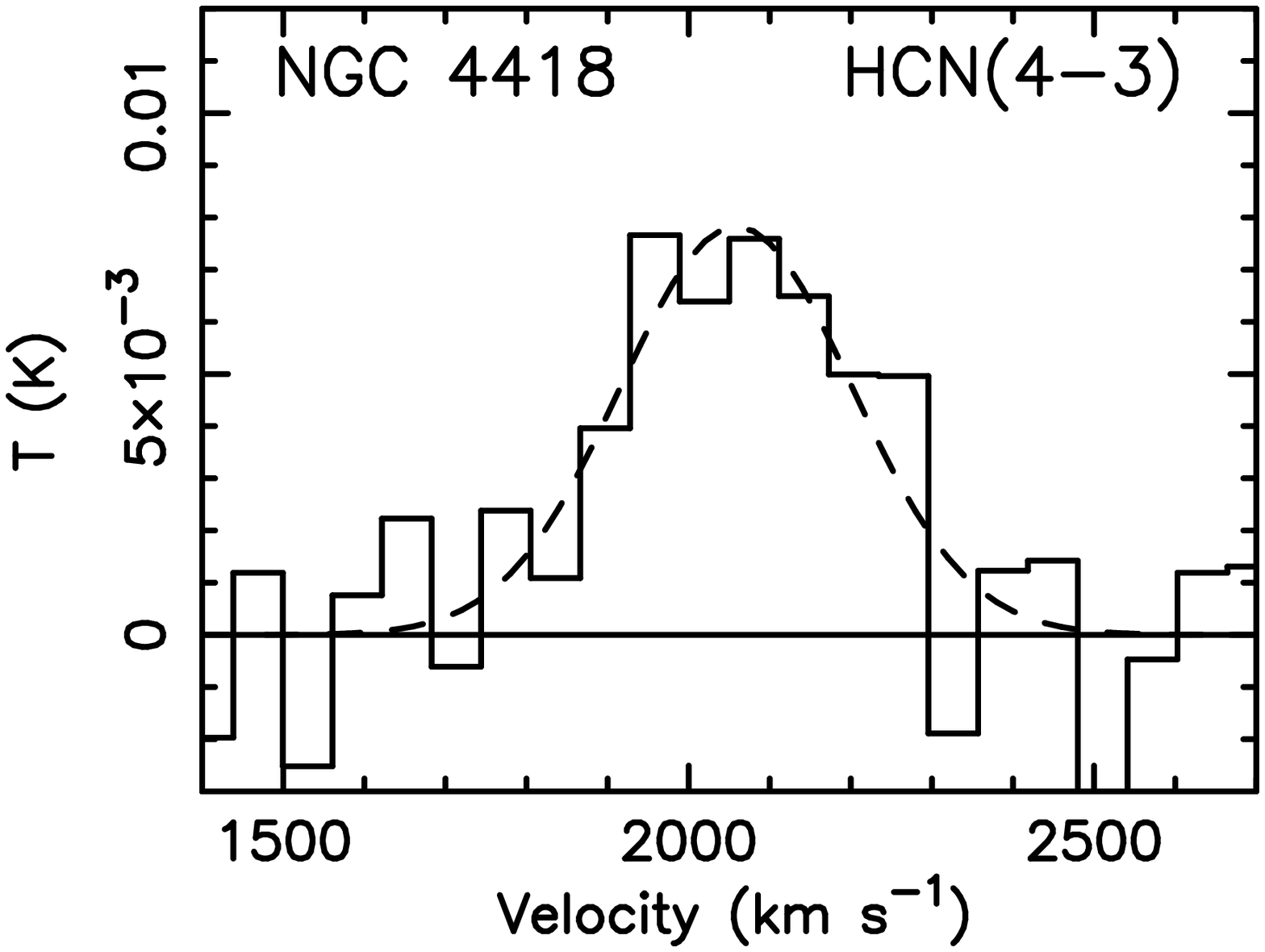}
\FigureFile(85mm,85mm){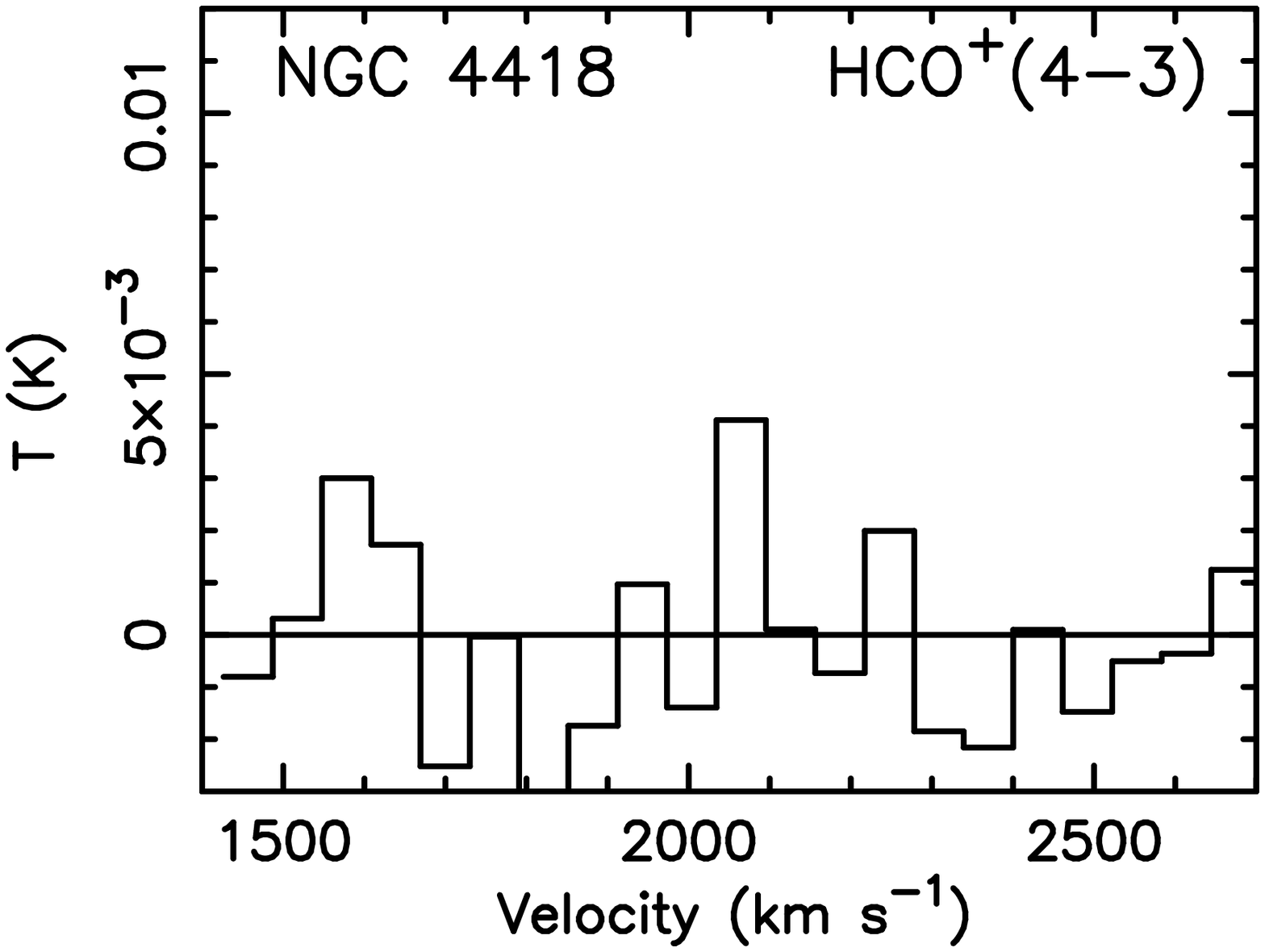}
\FigureFile(85mm,85mm){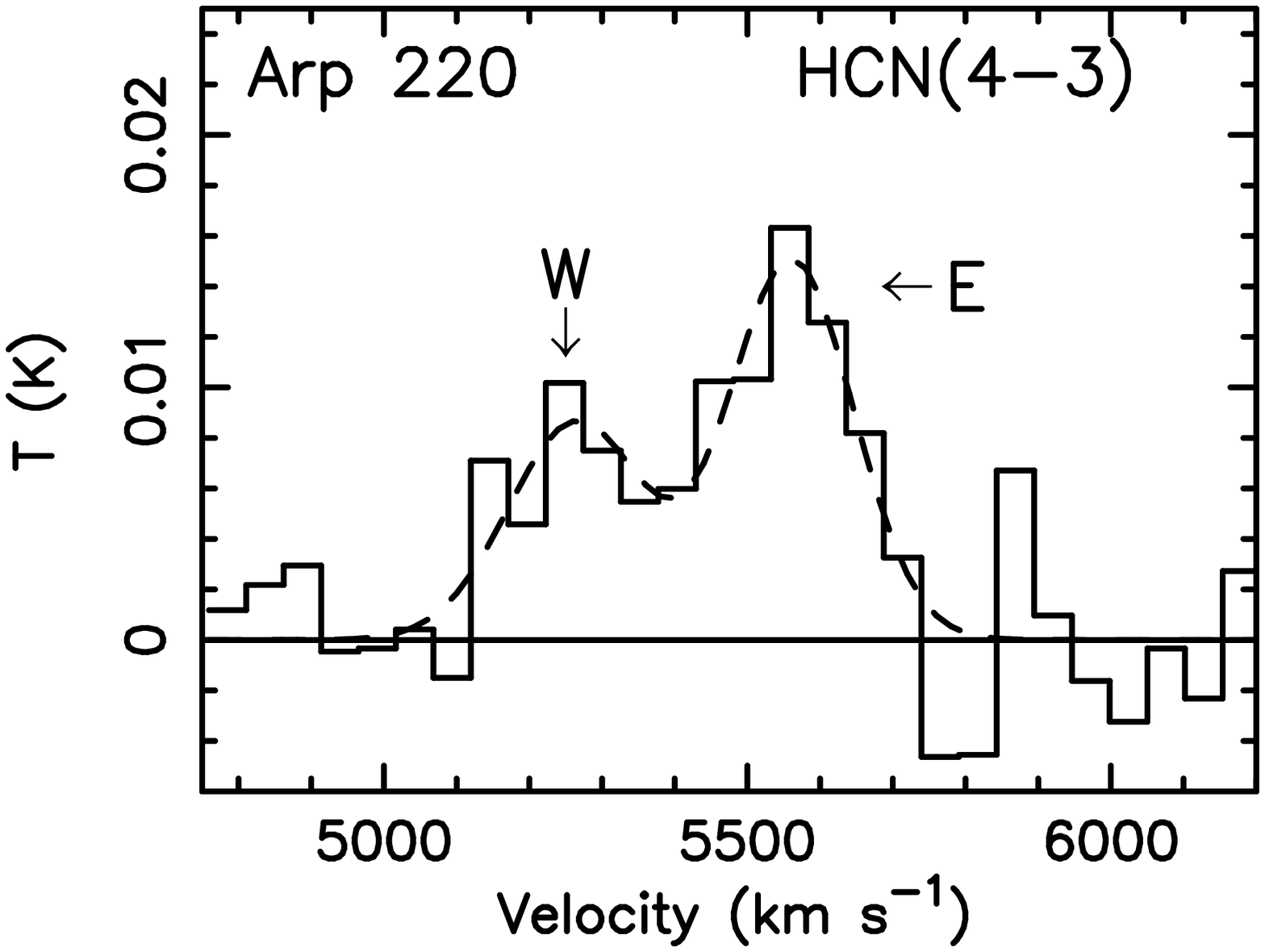}
\FigureFile(85mm,85mm){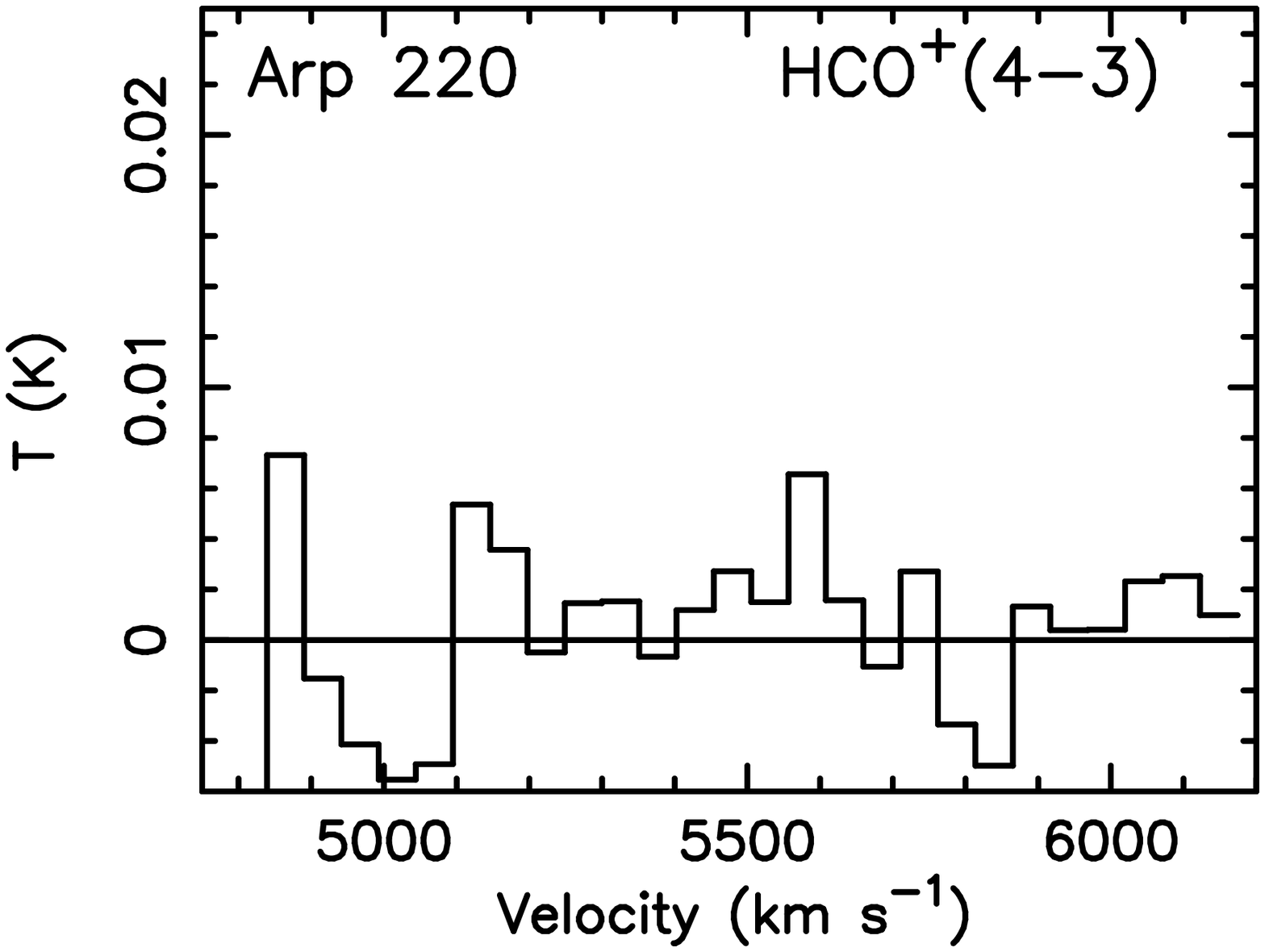}
%\FigureFile(80mm,80mm){ps/N4418HCNvel_v60Gauss.ps}
%\FigureFile(80mm,80mm){ps/N4418HCOvel_v60.ps}
%\FigureFile(80mm,80mm){ps/A220HCNvel_v50Gauss.ps}
%\FigureFile(80mm,80mm){ps/A220HCOvel_v50.ps}
\caption{
HCN(4--3) and HCO$^{+}$(4--3) spectra of NGC 4418 and Arp 220, after
baseline subtraction.
First-order linear fits were basically adopted as the baselines, except 
for the Arp 220 HCO$^{+}$(4--3) line, for which a second-order fit was
employed. 
The abscissa is velocity in [km s$^{-1}$] and the ordinate is antenna
temperature in [K].
Spectra of NGC 4418 and Arp 220 are smoothed to velocity resolutions
of 60 km s$^{-1}$ and 50 km s$^{-1}$, respectively.
Gaussian fits are over-plotted as dashed lines for the detected HCN(4--3)
lines in NGC 4418 and Arp 220.
For the Arp 220 HCN(4--3) line, we ascribe the blue (lower velocity) and
red (higher velocity) components to Arp 220 W (denoted as ``W'') and Arp
220 E (``E''), respectively, based on Nobeyama Millimeter Array  
interferometric HCN(1--0) and HCO$^{+}$(1--0) data \citep{ima07b} (see
$\S$4).  
The horizontal solid line indicates the zero flux level.
}
\end{figure}

%---- Figure 3 ----%
\begin{figure}
\FigureFile(85mm,85mm){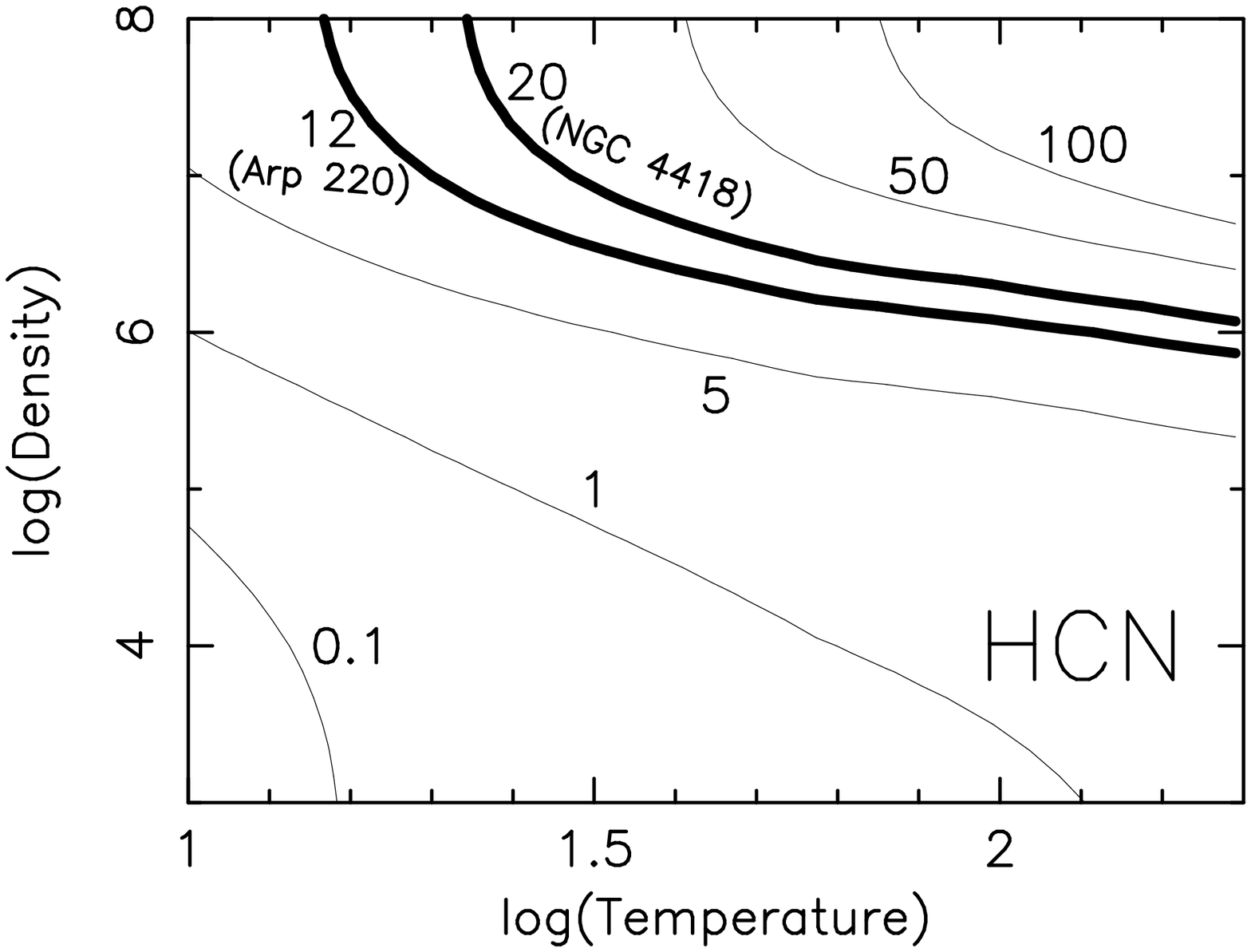} 
\FigureFile(85mm,85mm){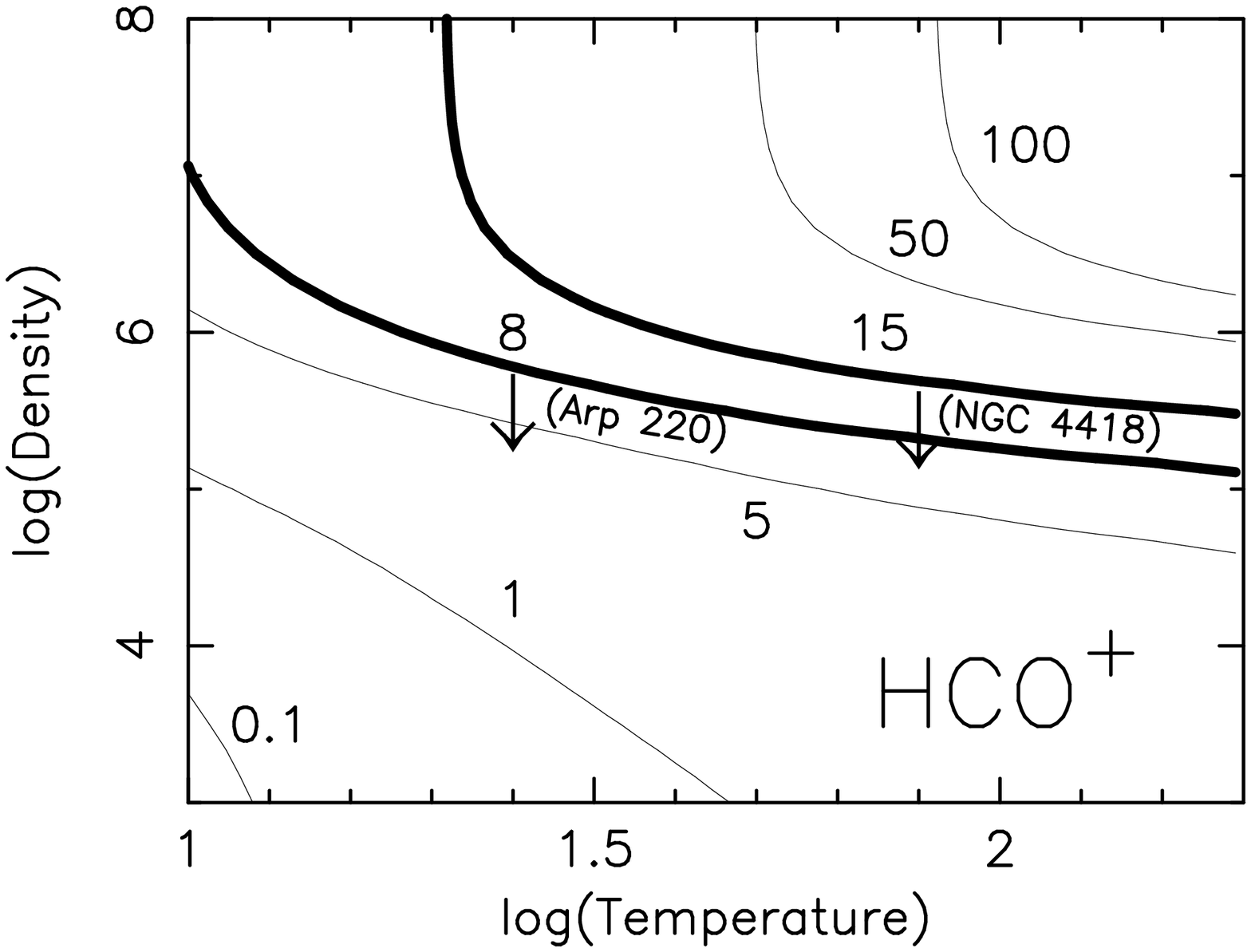} 
%\FigureFile(80mm,80mm){ps/HCN4-3_1-0.ps} 
%\FigureFile(80mm,80mm){ps/HCO+4-3_1-0.ps} 
\caption{
HCN(4--3) to HCN(1--0) ({\it Left}) and HCO$^{+}$(4--3) to
HCO$^{+}$(1--0) ({\it Right}) flux ratios in [Jy km s$^{-1}$] as a
function of gas number density and kinetic temperature, calculated with
RADEX.  
The abscissa and ordinate are the decimal logarithms of kinetic
temperature in [K] and H$_{2}$ gas number density in [cm$^{-3}$],
respectively. 
The HCN and HCO$^{+}$ column density is assumed to be 10$^{16}$
cm$^{-2}$ (see $\S$5.2). 
A velocity width of 200 km s$^{-1}$ was adopted (see $\S$5.2). 
Contours are 0.1, 1, 5, 12, 20, 50, and 100 for HCN, and 
0.1, 1, 5, 8, 15, 50, and 100 for HCO$^{+}$.
The measured ratios for NGC 4418 and Arp 220 are indicated as thick
lines. 
In the calculated parameter range, the line optical depth is 0--100
(i.e., reliable range in RADEX calculations).  
}
\end{figure}

%---- Figure 4 ----%
\begin{figure}
\FigureFile(85mm,85mm){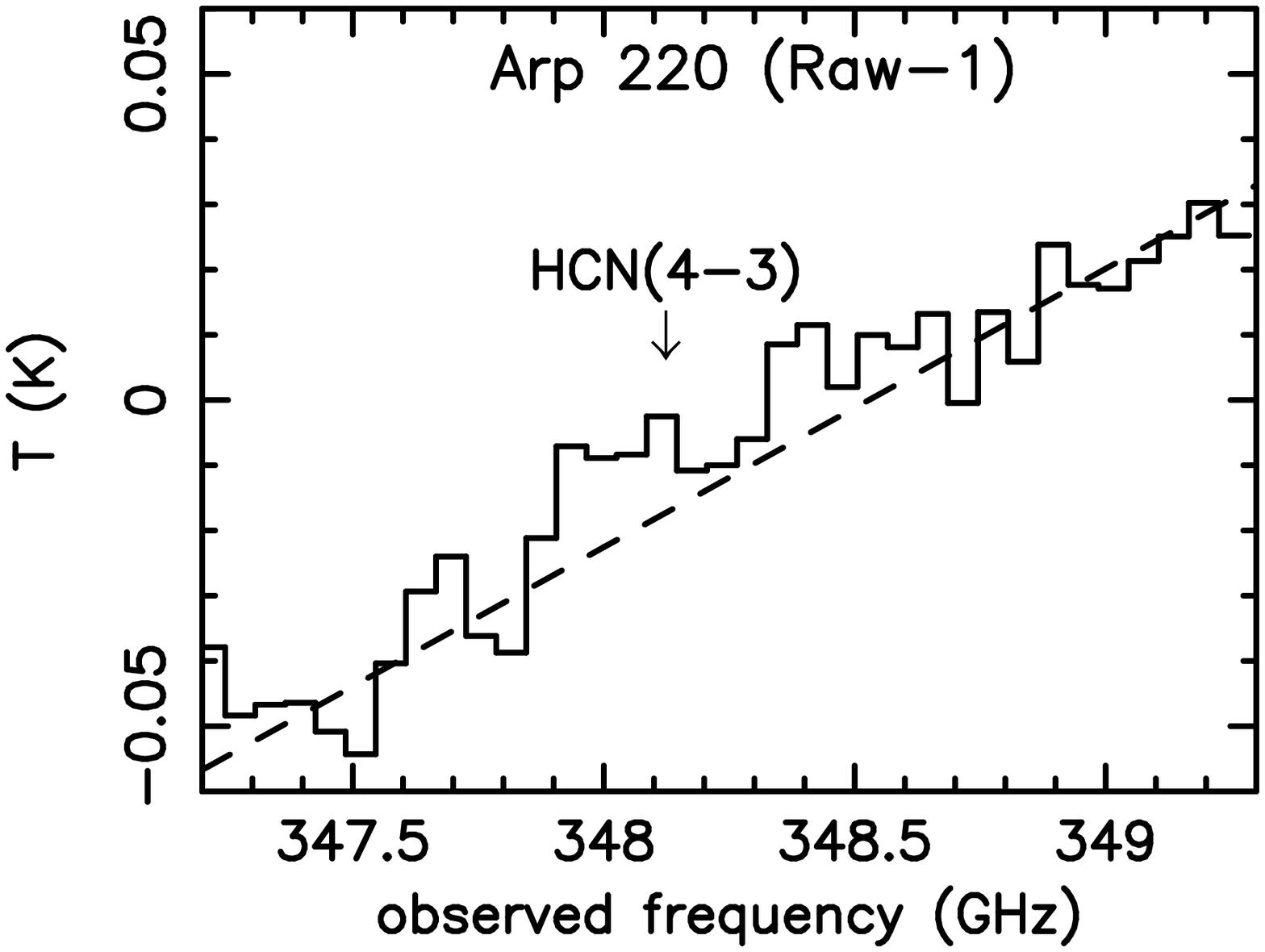} 
\FigureFile(85mm,85mm){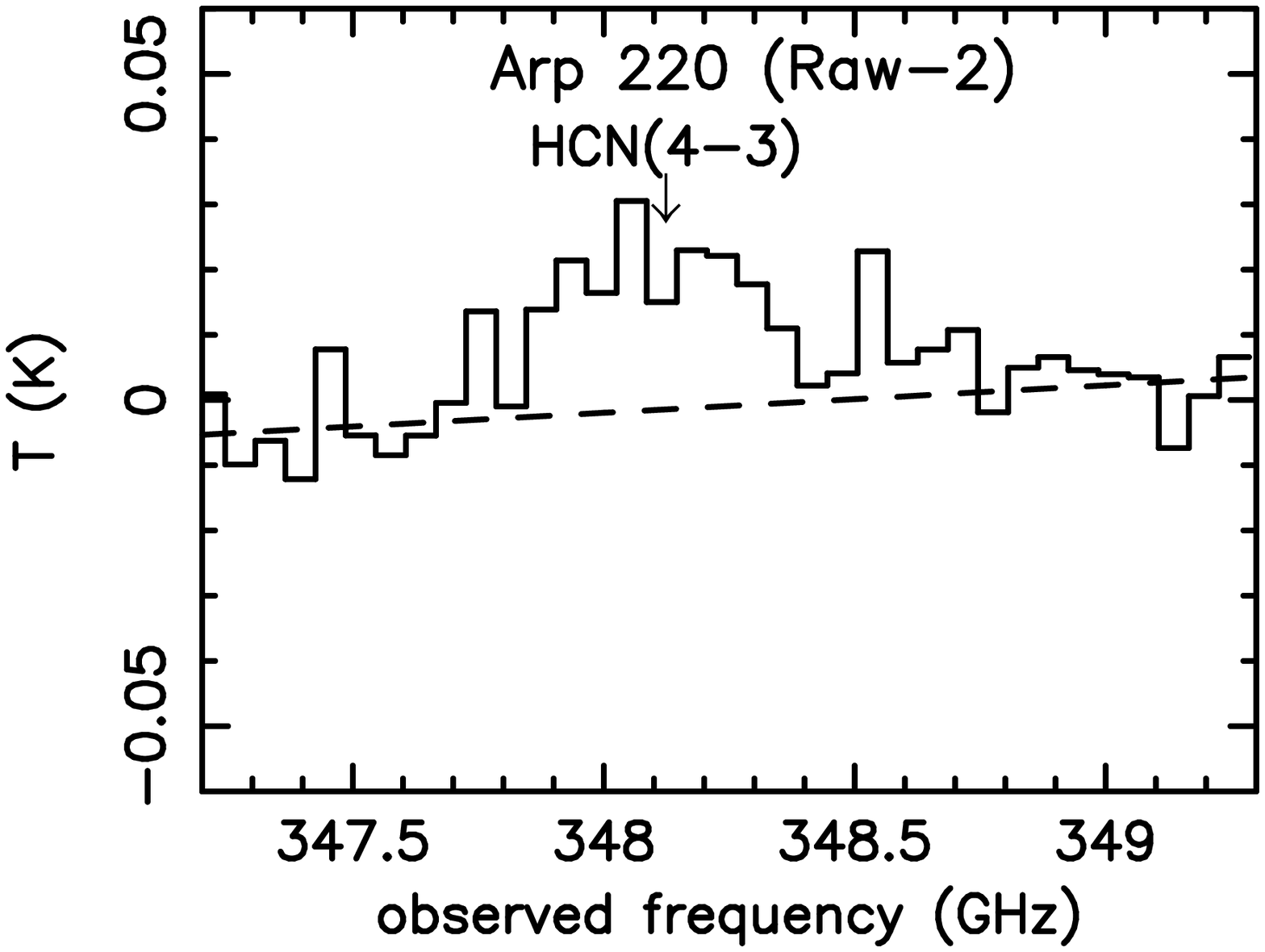} 
\FigureFile(85mm,85mm){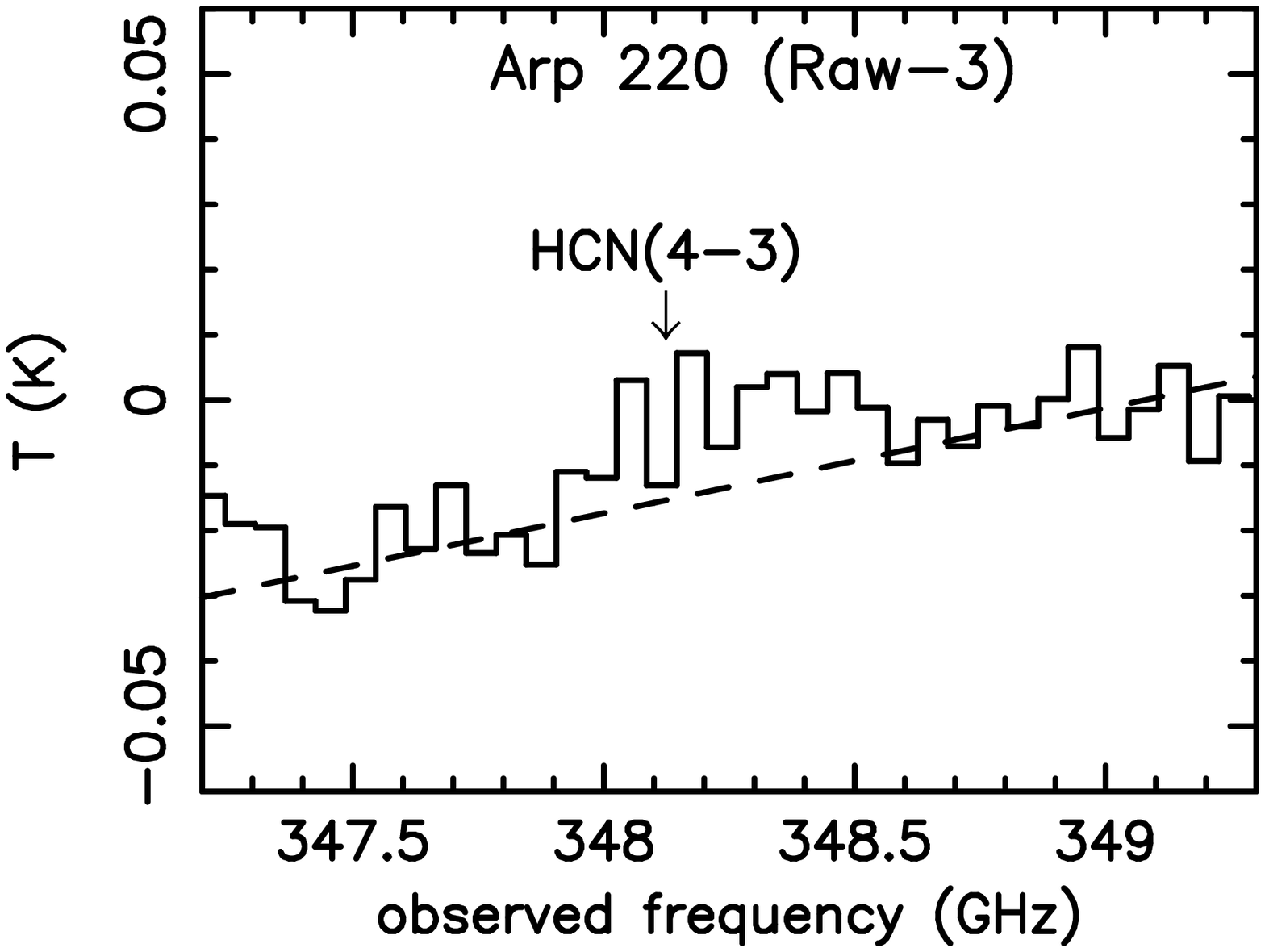} 
\FigureFile(85mm,85mm){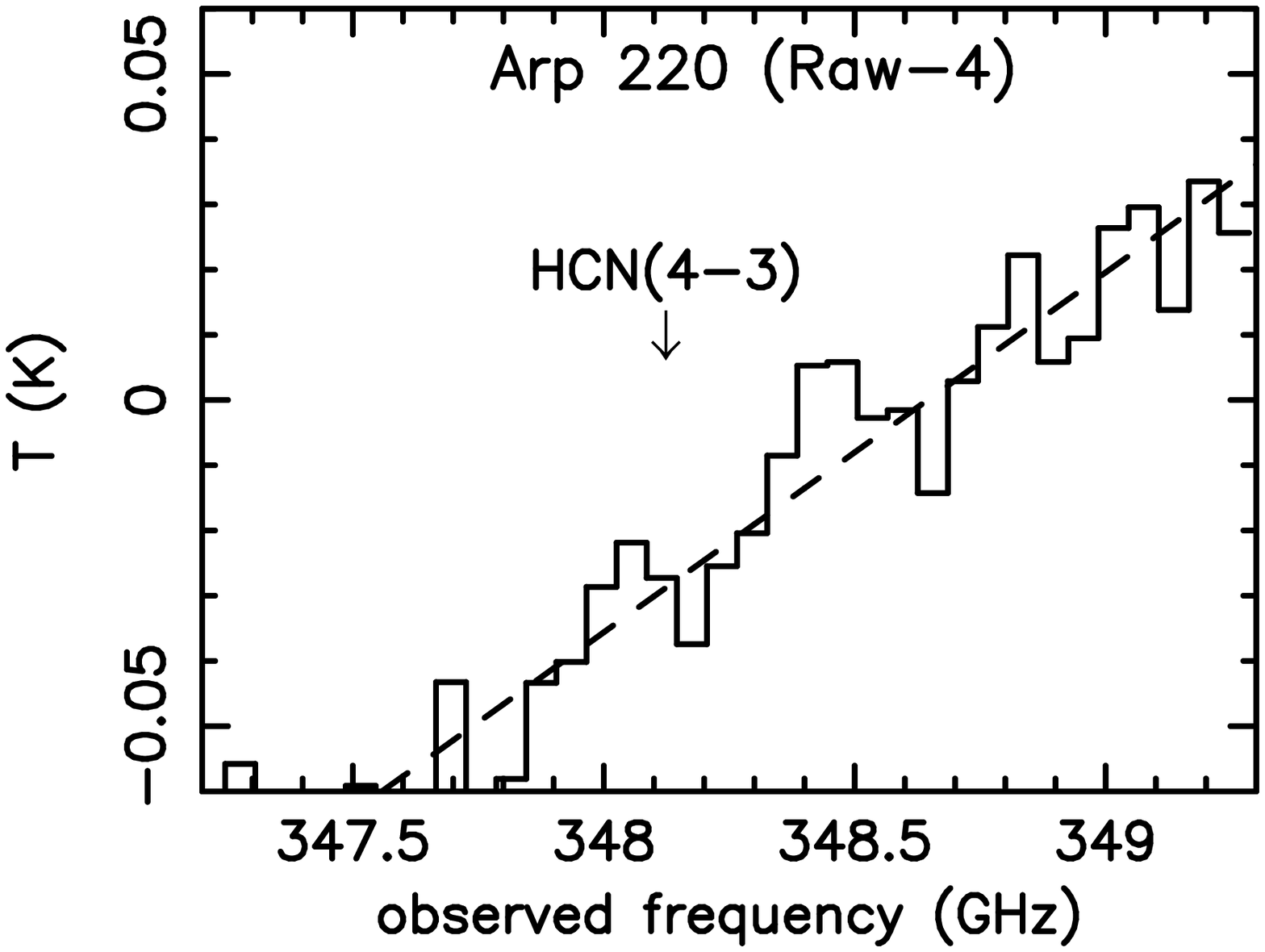} 
\FigureFile(85mm,85mm){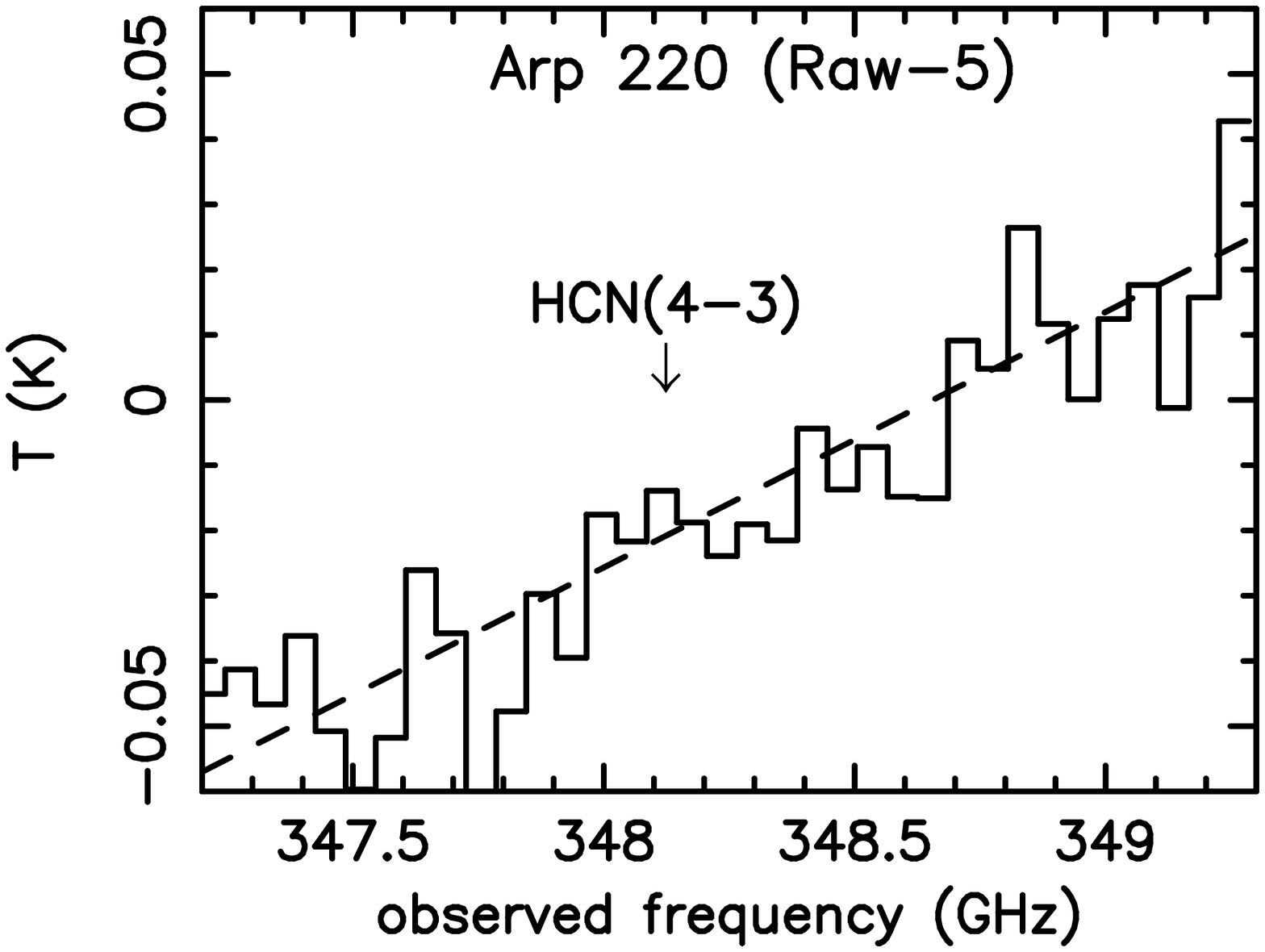} 
\FigureFile(85mm,85mm){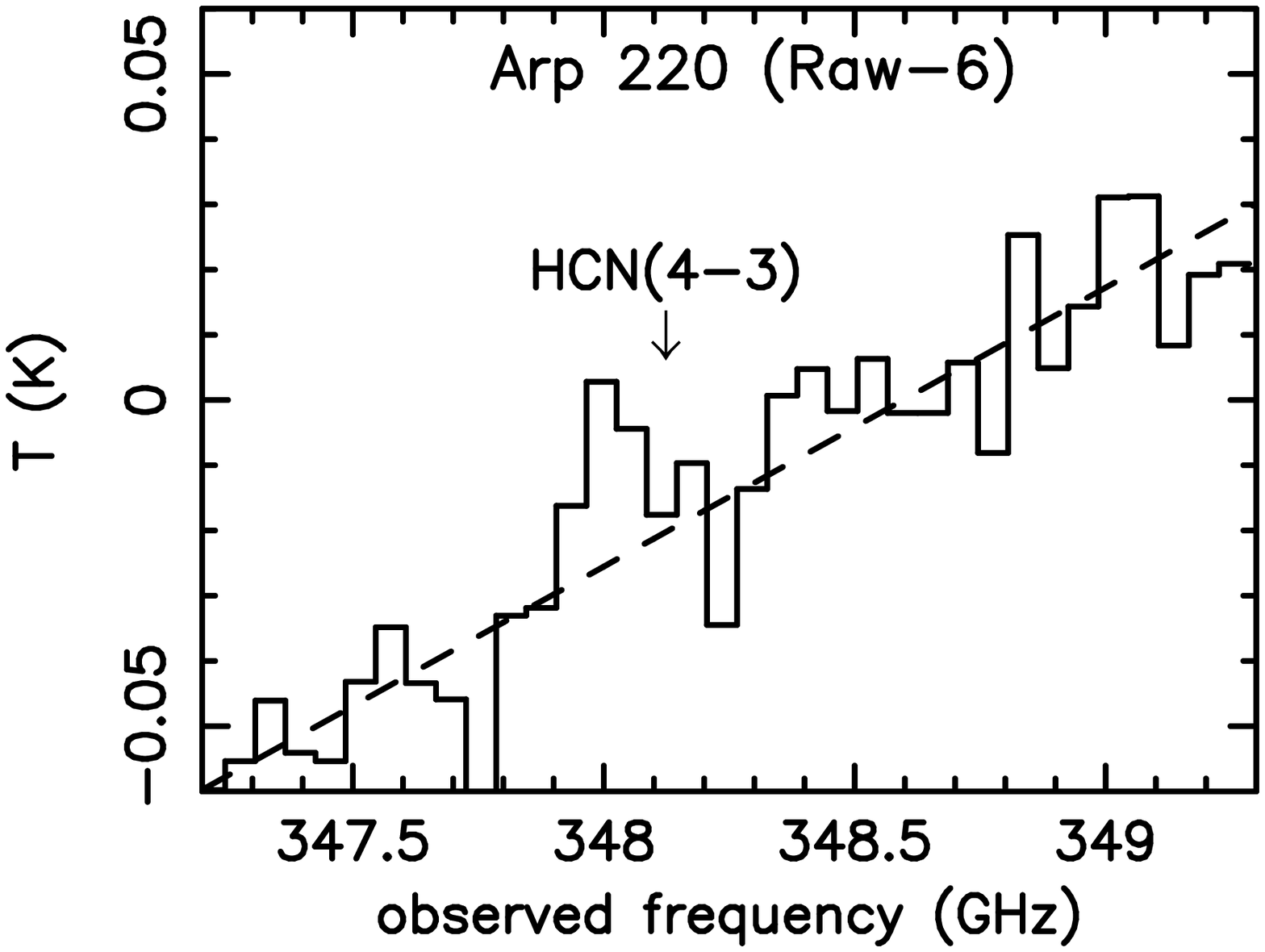} 
\end{figure}

\clearpage

\begin{figure}
\FigureFile(85mm,85mm){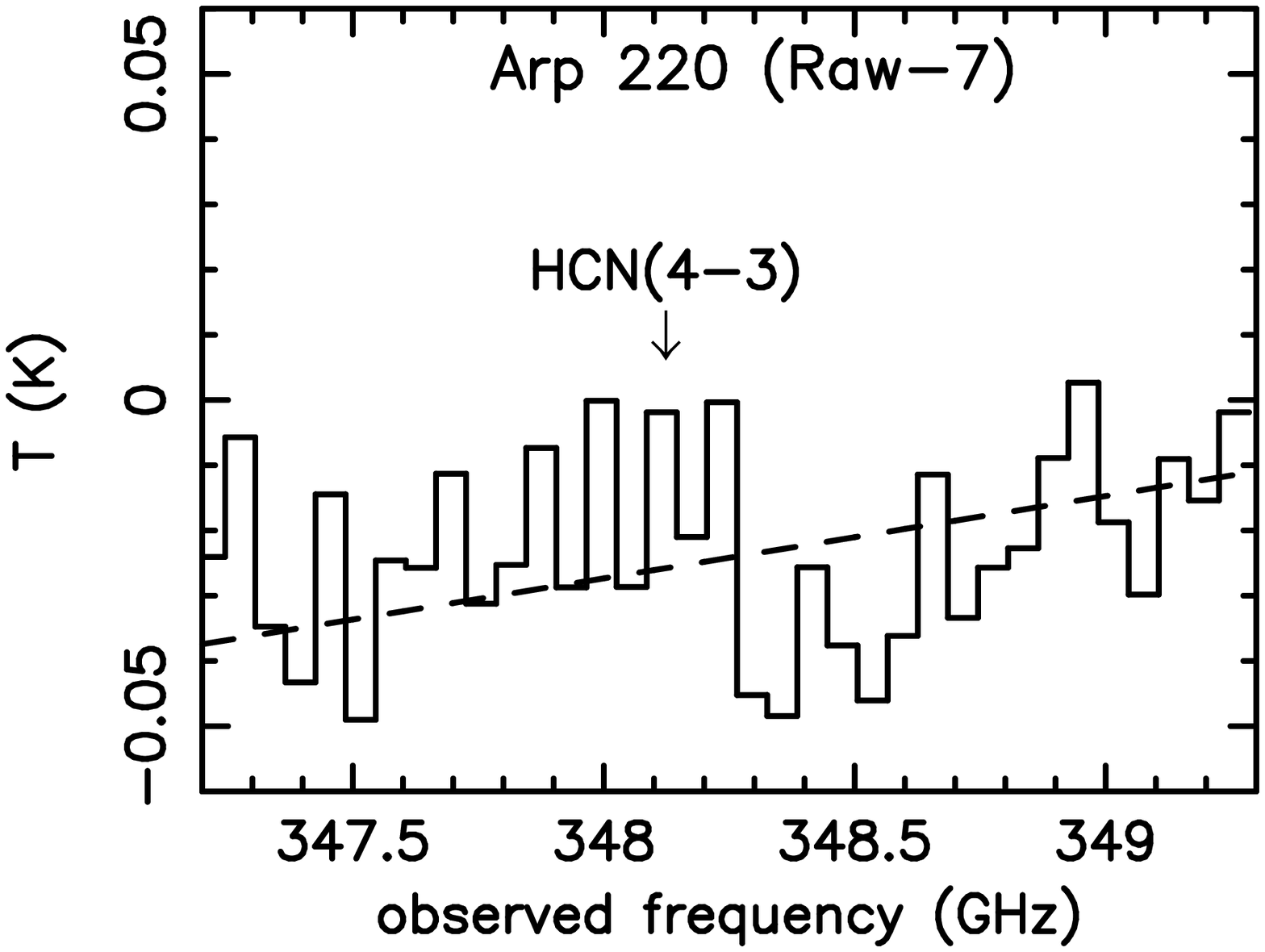} 
\FigureFile(85mm,85mm){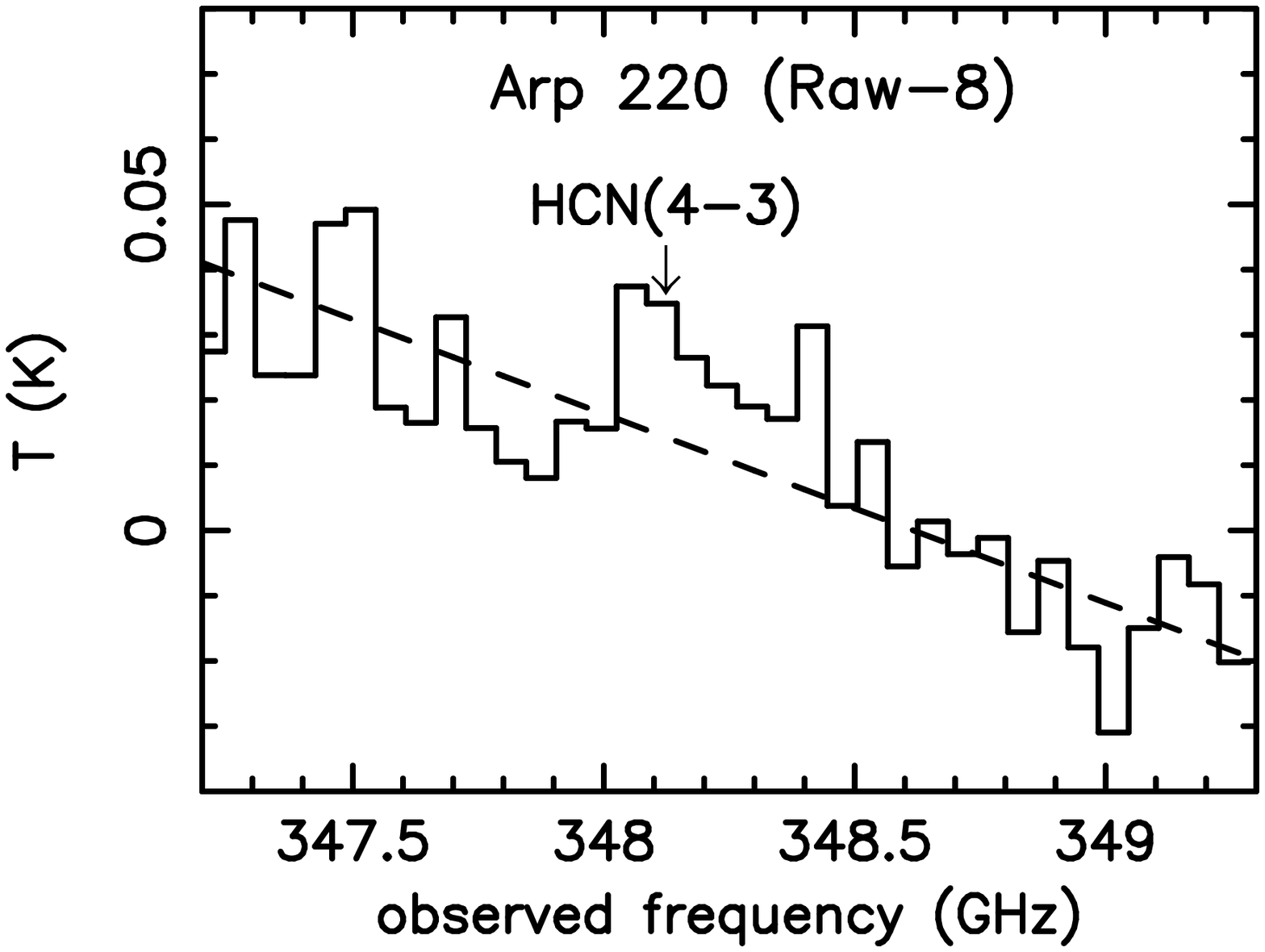} 
\caption{
Adopted linear baseline fits for individual unbinned HCN(4--3) spectra
of Arp 220 (dashed lines), and smoothed spectra with velocity
resolutions of 50 km s$^{-1}$ (solid lines). 
Signatures of the HCN(4--3) emission lines are seen in most of the
individual spectra. 
}
\end{figure}


\clearpage

\begin{thebibliography}{}
\bibitem[Aalto et al.(1995)]{aal95}
         Aalto, S., Booth, R. S., Black, J. H., \& Johansson,
         L. E. B. 1995, A\&A, 300, 369 
\bibitem[Aalto et al.(2007a)]{aal07a}
         Aalto, S., Spaans, M., Wiedner, M. C., \& Huttemeister, S. 
         2007a, A\&A, 464, 193
\bibitem[Aalto et al.(2007b)]{aal07b}
         Aalto, S., Monje, R., \& Martin, S. 2007b, A\&A, 475, 479
\bibitem[Armus, Heckman, \& Miley(1989)]{arm89} 
         Armus, L., Heckman, T. M., \& Miley, G. K. 1989, ApJ, 347, 727
\bibitem[Armus et al.(2007)]{arm07}
         Armus, L., et al. 2007, ApJ, 656, 148
\bibitem[Bardeen(1970)]{bar70}
         Bardeen, J. M. 1970, Nature, 226, 64
\bibitem[Blake et al.(1987)]{bla87}
         Blake, G. A., Sutton, E. C., \& Masson, C. R., \& Phillips,
         T. G. 1987, ApJ, 315, 621 
\bibitem[Dickens et al.(2000)]{dic00}
         Dickens, J. E., Irvine, W. M., Snell, R. L., Bergin, E. A., 
         Schloerb, F. P., Pratap, P., \& Miralles, M. P. 2000, ApJ, 542,
         870
\bibitem[Downes \& Eckart(2007)]{dow07}
         Downes, D., \& Eckart, A. 2007, A\&A, 468, L57
\bibitem[Dudley \& Wynn-Williams(1997)]{dud97} 
         Dudley, C. C., \& Wynn-Williams, C. G. 1997, ApJ, 488, 720 
\bibitem[Ezawa et al.(2004)]{eza04}
         Ezawa, H. et al. 2004, Proc. SPIE, 5489, 763 
\bibitem[Ezawa et al.(2008)]{eza08}
         Ezawa, H. et al. 2008, Proc. SPIE, 7012, 6
\bibitem[Garcia-Burillo et al.(2006)]{gar06}
         Garcia-Burillo, S., et al. 2006, ApJ, 645, L17 
\bibitem[Gonzalez-Alfonso et al.(2004)]{gon04} 
         Gonzalez-Alfonso, E., Smith, H. A., Fischer, J., \& Cernicharo,
         J., 2004, ApJ, 613, 247
\bibitem[Gracia-Carpio et al.(2008)]{gra08} 
         Gracia-Carpio, J., Garcia-Burillo, S, Planesas, P., 
         Fuente, A., \& Usero, A. 2008, A\&A, 479, 703
\bibitem[Greve et al.(2009)]{gre09}
         Greve, T. R., Papadopoulos, P. P., Gao, Y., \& Radford,
         S. J. E. 2009, ApJ, 692, 1432 
\bibitem[Guelin et al.(2007)]{gue07} 
         Guelin, M., et al. 2007, A\&A, 462, L45
\bibitem[Iguchi \& Okuda(2008)]{igu08}
         Iguchi, S., \& Okuda, T. 2008, PASJ, 60, 857  
\bibitem[Imanishi et al.(2006a)]{idm06} 
         Imanishi, M., Dudley, C. C., \& Maloney, P. R. 2006a, ApJ,
         637, 114
\bibitem[Imanishi et al.(2007a)]{ima07a} 
         Imanishi, M., Dudley, C. C., Maiolino, R., Maloney, P. R., 
         Nakagawa, T., \& Risaliti, G. 2007a, ApJS, 171, 72
\bibitem[Imanishi \& Nakanishi(2006)]{in06} 
         Imanishi, M., \& Nakanishi, K. 2006, PASJ, 58, 813
\bibitem[Imanishi et al.(2004)]{ima04}
         Imanishi, M., Nakanishi, K., Kuno, N., \& Kohno, K. 2004, AJ,
         128, 2037 
\bibitem[Imanishi et al.(2006b)]{ink06} 
         Imanishi, M., Nakanishi, K., \& Kohno, K. 2006b, AJ, 131, 2888
\bibitem[Imanishi et al.(2007b)]{ima07b} 
         Imanishi, M., Nakanishi, K., Tamura, Y., Oi, N., \& Kohno,
         K. 2007b, AJ, 134, 2366
\bibitem[Imanishi et al.(2009)]{ima09} 
         Imanishi, M., Nakanishi, K., Tamura, Y., \& Peng, C. -H. 2009,
         AJ, 137, 3581 
\bibitem[Iwasawa et al.(2005)]{iwa05} 
         Iwasawa, K., Sanders, D. B., Evans, A. S., Trentham, N.,
         Miniutti, G., \& Spoon, H. W. W. 2005, MNRAS, 357, 565 
\bibitem[Kamazaki et al.(2005)]{kam05}
         Kamazaki, T., et al. 2005, Astronomical Society of the Pacific 
         Conference Series, 347, 533 
\bibitem[Kohno(2005)]{koh05}
         Kohno, K. 2005, in AIP Conf. Ser. 783, 
         The Evolution of Starbursts, ed. S. H\"uttemeister, E. Manthey,
         D. Bomans, \& K. Weis (New York: AIP), 203 (astro-ph/0508420)
\bibitem[Krips et al.(2008)]{kri08}
         Krips, M., Neri, R., Garcia-Burillo, S., Martin, S., Combes,
         F., Gracia-Carpio, J., \& Eckart, A. 2008, ApJ, 677, 262 
\bibitem[Lahuis et al.(2007)]{lah07}
         Lahuis, F. et al. 2007, ApJ, 659, 296 
\bibitem[Lehnert \& Heckman(1995)]{leh95}
         Lehnert, M. D., \& Heckman, T. M. 1995, ApJS, 97, 89
\bibitem[Maiolino et al.(2003)]{mai03}
         Maiolino, R. et al. 2003, MNRAS, 344, L59
\bibitem[Meijerink \& Spaans(2005)]{mei05}
         Meijerink, R., \& Spaans, M. 2005, A\&A, 436, 397 
\bibitem[Okuda \& Iguchi(2008)]{oku08}
         Okuda, T., \& Iguchi, S. 2008, PASJ, 60, 315
\bibitem[Pratap et al.(1997)]{pra97}
         Pratap, P., Dickens, J. E., Snell, R. L., Miralles, M. P.,
         Bergin, E. A., Irvine, W. M., \& Schloerb, F. P. 1997, ApJ,
         486, 862  
\bibitem[Riechers et al.(2006)]{rie06}
         Riechers, D. et al. 2006, ApJ, 650, 604
\bibitem[Roche et al.(1991)]{roc91}
         Roche, P. F., Aitken, D. K., Smith, C. H., \& Ward, M. J. 
         1991, MNRAS, 248, 606 
\bibitem[Sakamoto et al.(2008)]{sak08}
         Sakamoto, K. et al. 2008, ApJ, 684, 957   
\bibitem[Sakamoto et al.(2009)]{sak09}
         Sakamoto, K. et al. 2009, ApJ, 700, L104
\bibitem[Sanders \& Mirabel(1996)]{sam96}
         Sanders, D. B., \& Mirabel, I. F. 1996, ARA\&A, 34, 749
\bibitem[Scoville et al.(2000)]{sco00}
         Scoville, N. Z. et al. 2000, AJ, 119, 991  
\bibitem[Soifer et al.(2000)]{soi00}
         Soifer, B. T. et al. 2000, AJ, 119, 509 
\bibitem[Spoon et al.(2001)]{spo01} 
         Spoon, H. W. W., Keane, J. V., Tielens, A. G. G. M., Lutz, D.,
         \& Moorwood, A. F. M. 2001, A\&A, 365, L353 
\bibitem[Spoon et al.(2004)]{spo04}
         Spoon, H. W. W., Moorwood, A. F. M., Lutz, D., Tielens,
         A. G. G. M., Siebenmorgen, R., \& Keane, J. V. 2004, A\&A, 414,
         873  
\bibitem[Thompson et al.(2005)]{tho05}
         Thompson, T. A., Quataert, E., \& Murray, N. 2005, ApJ, 
         630, 167
\bibitem[Thorne(1974)]{tho74}
         Thorne, K. S. 1974, ApJ, 191, 507
\bibitem[van der Tak et al.(2007)]{van07}
         van der Tak, F. F. S., Black, J. H., Schoier, F. L., Jansen,
         D. J., \& van Dishoeck, E. F. 2007, A\&A, 468, 627  
\bibitem[Veilleux et al.(1999)]{vei99} 
         Veilleux, S., Kim, D. -C., \& Sanders, D. B. 1999, ApJ, 
         522, 113
\bibitem[Veilleux \& Osterbrock(1987)]{vei87} 
         Veilleux, S., \& Osterbrock, D. E. 1987, ApJS, 63, 295 
\bibitem[Veilleux et al.(2009)]{vei09} 
         Veilleux, S., et al. 2009, ApJS, 182, 628
\bibitem[Wada \& Norman(2007)]{wad07}
         Wada, K., \& Norman, C. 2007, ApJ, 660, 276
\bibitem[Wang et al.(1994)]{wan94}
         Wang, Y., Jaffe, D. T., Graf, U. U., \& Evans II, N. J. 1994, 
         ApJS, 95, 503
\bibitem[Weiss et al.(2007)]{wei07}
         Weiss, A., Downes, D., Neri, R., Walter, F., Henkel, C., 
         Wilner, D. J., Wagg, J., \& Wiklind, T. 2007, A\&A, 467, 955  
\bibitem[Werner et al.(1976)]{wer76}
         Werner, M. W., Gatley, I., Harper, D. A., Becklin, E. E., 
         Loewenstein, R. F., Telesco, C. M., \& Thronson, H. A. 1976,
         ApJ, 204, 420 
\bibitem[Wiedner et al.(2002)]{wie02}
         Wiedner, M. C., Wilson, C. D., Harrison, A., Hills, R. E., Lay,
         O. P., \& Carlstrom, J. E., 2002, ApJ, 581, 299
\bibitem[Yamada et al.(2007)]{yam07}
         Yamada, M., Wada, K., \& Tomisaka, K. 2007, ApJ, 671, 73 
\end{thebibliography}
\end{document}